\documentclass[aps,prb,reprint,groupedaddress, showpacs]{revtex4-1}
\usepackage{graphicx,graphics}
\usepackage{epstopdf} 
\usepackage{amsmath,amssymb,amsfonts}
\usepackage{latexsym,verbatim}
\usepackage{bm}
\usepackage{bbold}
\usepackage{color}
\usepackage[normalem]{ulem}
\usepackage[breaklinks=false,colorlinks,citecolor=blue,linkcolor=blue,urlcolor=blue]{hyperref}
\usepackage[abs]{overpic}
\usepackage{textcomp} 
\usepackage{mathtools}
\usepackage{braket}
\usepackage{soul}    

\begin{document}

\title{Interacting holes in Si and Ge double quantum dots: from a multiband approach to an effective--spin picture}

\author{Andrea Secchi}
\email{andrea.secchi@nano.cnr.it}
\author{Laura Bellentani}
\author{Andrea Bertoni}
\author{Filippo Troiani}
\affiliation{Centro S3, CNR-Istituto di Nanoscienze, I-41125 Modena, Italy}

\begin{abstract}

The states of two electrons in tunnel--coupled semiconductor quantum dots can be effectively described in terms of a two--spin Hamiltonian with an isotropic Heisenberg interaction. A similar description needs to be generalized in the case of holes due to their multiband character and spin--orbit coupling, which mixes orbital and spin degrees of freedom, and splits $J=3/2$ and $J = 1/2$ multiplets. Here we investigate two--hole states in prototypical coupled Si and Ge quantum dots via different theoretical approaches. Multiband $\boldsymbol{k}\cdot\boldsymbol{p}$ and Configuration--Interaction calculations are combined with entanglement measures in order to thoroughly characterize the two--hole states in terms of band mixing and justify the introduction of an effective spin representation, which we analytically derive a from generalized Hubbard model. We find that, in the weak interdot regime, the ground state and first excited multiplet of the two--hole system display -- unlike their electronic counterparts -- a high degree of $J$--mixing, even in the limit of purely heavy--hole states. The light--hole component additionally induces $M$--mixing and a weak coupling between spinors characterized by different permutational symmetries. 

\end{abstract}

\date{\today}

\maketitle

\section{Introduction}

Semiconductor quantum dots (QDs) and double quantum dots (DQDs) \cite{Burkard99a, Climente2008a, Yakimov2010a} represent the solid--state analogues of atoms and biatomic molecules, respectively. Unlike in the case of atoms, the main QD features, such as the shape of the confining potential and the interdot coupling, can be widely tuned by means of electrostatic gates. Besides, the properties of confined particles in QDs are affected and diversified by the host semiconductor and its band structure.

From a technological viewpoint, one of the most relevant applications of QDs is the implementation of spin qubits\cite{Loss98}. In this perspective, Si/Ge nanostructures seem to represent ideal building blocks. In fact, both semiconductors present a reduced hyperfine interaction, due to the natural abundance of nonmagnetic isotopes ($>95\%$ and $>92\%$ for Si and Ge, respectively) and to the possibility of isotopically purifying the samples\cite{Zwanenburg2013, Scappucci2020}. Moreover, the availability of well--established industrial technologies for the production of Si/Ge nanostructures, such as metal--oxide--semiconductor devices, is possibly a crucial asset towards scalability\cite{Horibe2015, Yamaoka2016}. 

Although a great deal of work has been performed on conduction--band Si QDs\cite{Hada2004, Wang2010, Wang2011, DasSarma2011, Simmons2011, Raith2012, Nielsen2012, Watson2018, Zajac2018, Ansaloni2020}, the implementation of hole QDs has recently attracted considerable interest for quantum applications \cite{Maurand2016, Bonen2019, Hetenyi2020a, Watzinger2016, Watzinger2018, Hendrickx2019, Hendrickx2020, Scappucci2020, Terrazos2021, Bosco2021}. One reason is that the valence band of both Si and Ge does not display the 6--fold valley degeneracy that characterizes the conduction band\cite{Hada2003} and provides an unwanted additional degree of freedom. Furthermore, the valence band is generated by the hybridization of $p$ atomic orbitals \cite{Voon2009}, which have nodes at the atomic nuclei, so that the residual hyperfine interaction affects the hole states only weakly. In the case of hole--spin qubits, germanium presents a further additional advantage \cite{Watzinger2016, Watzinger2018, Hendrickx2019, Hendrickx2020, Scappucci2020, Terrazos2021}: because of their small effective mass, Ge holes tunnel more efficiently between neighboring QDs, which implies less stringent requirements on dot sizes and interdot distances.

Si and Ge have six valence bands with a maximum at the $\boldsymbol{\Gamma}$ point, which, at zero magnetic field, form three Kramers--degenerate couples which are respectively called heavy--hole, light--hole, and split--off bands \cite{Chao92, Voon2009}. The heavy-- and light-- hole bands are degenerate at $\boldsymbol{\Gamma}$, while the split--off bands lie at a higher energy that is equal to the spin--orbit parameter. This scenario results from a strong spin--orbit coupling between the hole spins and orbital angular momenta, which can be exploited for efficient spin manipulation through electric--dipole spin resonance\cite{Bulaev07, Watzinger2018}. The much weaker spin--orbit coupling acting in the conduction band makes the implementation of this concept more challenging for electrons \cite{Corna2018}.

The existence of distinct valence bands also complicates the character of two--particle states in coupled quantum dots. In a single--band system, interdot tunneling induces a hybridization of the single--dot orbitals. In the case of weak tunneling and identical dots, the single--particle spatial wave functions can be approximately identified with symmetric and antisymmetric combinations of single--dot orbitals. The lowest two--particle eigenstates result from the interplay between tunneling and Coulomb interactions and, for weak spin--orbit coupling, they can be assigned well--defined values of the total spin $S$. In particular, the ground and first excited states respectively correspond to a spin singlet ($S=0$) and triplet ($S=1$) \cite{Cerfontaine2020}.

In a multiband system, instead, the single--dot ground state is a combination of different Bloch states corresponding to the distinct bands, each of which might be coupled to a different orbital envelope function. For tunnel--coupled quantum dots, this results in a richer and more complicated picture than the one outlined in the single--band case. The lowest molecular orbitals are determined by the competition between different excitations, associated to interdot and band degrees of freedom. In particular geometries, the interplay between confinement potential and spin--orbit coupling can lead to the appearance of anomalous features, like the vanishing of the tunneling energy at finite interdot distances, and the related vanishing of the singlet--triplet splitting \cite{Climente2008a, Doty2008, Yakimov2012, Deng2018a}.

The purpose of this work is to elucidate the effect of the multiband structure of the valence band on the single-- and two--hole wave functions of Si and Ge confined states. We consider electrostatically--defined DQDs, where tunneling takes place within a given quantum well and across a smooth barrier, modulated by top gates. The single-- and two--hole states are determined for a model DQD potential within a six--band $\boldsymbol{k} \cdot \boldsymbol{p}$ envelope--function approach combined with the Configuration--Interaction (CI) scheme for the diagonalization of the two--hole interacting Hamiltonian. The numerical results are taken as the benchmark and the starting point for an effective representation in terms of an analytical model. The latter is inspired by the Hubbard model, with the two QDs playing the role of the atomic sites and two Kramers--degenerate single--hole spin--orbital states for each QD. Crucially, the band structure of each of these spin--orbitals allows us to derive analytical expressions for both the two--hole eigenstates and the reduced spin states. We find that the two--hole states are non--trivial mixtures of spinors with different values of the total spin $J$ and its third component $M$. As a central result, the predictions of the Hubbard model are checked against the numerical modeling and reveal a very good agreement.  

Our second main result follows from the numerical calculations of the linear entropies associated with the reduced spin--density matrices of the lowest eigenstates of the single-- and two--hole systems. These calculations reveal that spin--orbit entanglement is rather weak in these systems; this implies that the single--hole orbital wave functions corresponding to different bands are approximately parallel (i.e., they differ by a multiplicative constant). This allows us to simplify the results of the Hubbard model, and to derive a simpler description of single-- and two--hole states in terms of pseudospin--$1/2$ states. This demonstrates that linear entropies are a powerful tool for the analysis and the derivation of effective models for multiband systems.

The rest of this article is organized as follows. In Sections \ref{Sec: Single} and \ref{Sec: Two} we present the calculation and characterization of single-- and two--hole states, respectively. Section \ref{Sec:Hubbard} is devoted to the analytical Hubbard model and to the comparison between its predictions and the numerical results. In Sec.~\ref{Sec: spin rep} we introduce the approximation of neglecting spin--orbit entanglement, which allows to obtain from the Hubbard model a simpler pseudospin--$1/2$ representation of single-- and two--hole states. Finally, the conclusions are drawn in Sec.~\ref{Sec: Conclusions}. Further technical details related to the comparison between the numerical results and the 4--band Hubbard model are reported in Appendix \ref{App: numerical 2 holes vs Hubbard}.

\section{Single--hole states}
\label{Sec: Single}

\subsection{Method}

\subsubsection{Diagonalization of the single--hole Hamiltonian}

The calculation of the confined single--hole states in Si and Ge is performed within the L\"uttinger--Kohn envelope--function approach \cite{Luttinger55}. As a first step, the kinetic--energy operator for the electronic states close to the top of the valence bands (which occurs at the $\boldsymbol{\Gamma}$ point in both Si and Ge) is represented by a $\boldsymbol{k}\cdot \boldsymbol{p}$ Hamiltonian matrix ($H_{\boldsymbol{k} \cdot \boldsymbol{p}}$). This acts on vectors, each component of which corresponds to a Bloch state with crystal momentum $\boldsymbol{\Gamma} \equiv \boldsymbol{0}$. In the cases of Si and Ge, the relevant Bloch states for the valence bands are built from $p$--type atomic orbitals \cite{Voon2009, Secchi2020}, carrying an angular momentum $l=1$. Combining this with the electron $s=1/2$ spin, one can write the Bloch basis set at $\boldsymbol{\Gamma}$ as a quartet of states with $j=3/2$ and a doublet of states with $j=1/2$. The quartet generates the heavy-- ($m=\pm 3/2$) and light--hole ($m=\pm 1/2$) bands, while the doublet generates the spin--orbit split--off bands. The $\boldsymbol{k} \cdot \boldsymbol{p}$ Hamiltonian is therefore a $6 \times 6$ matrix which, in the $\left\{ \left| \frac{3}{2}, \frac{3}{2} \right>, \left| \frac{3}{2}, \frac{1}{2} \right>, \left| \frac{3}{2}, -\frac{1}{2} \right>, \left| \frac{3}{2}, -\frac{3}{2} \right>, \left| \frac{1}{2}, \frac{1}{2} \right>, \left| \frac{1}{2}, -\frac{1}{2} \right> \right\}$ basis, reads as \cite{Voon2009, Chao92} 
\begin{align}
    H_{\boldsymbol{k} \cdot \boldsymbol{p}} = 
    \left(
    \begin{matrix}
    P+Q  & -S & R & 0 & - \frac{1}{\sqrt{2}} S & \sqrt{2} R \\
    -S^* & P-Q & 0 & R & - \sqrt{2} Q & \sqrt{\frac{3}{2}} S \\
    R^* & 0 & P-Q & S & \sqrt{\frac{3}{2}} S^* & \sqrt{2} Q \\
    0 & R^* & S^* & P+Q & - \sqrt{2} R^* & - \frac{1}{\sqrt{2}} S^* \\
    - \frac{1}{\sqrt{2}} S^* & - \sqrt{2} Q & \sqrt{\frac{3}{2}} S & - \sqrt{2} R & P + \Delta & 0 \\
    \sqrt{2} R^* & \sqrt{\frac{3}{2}} S^* & \sqrt{2} Q & - \frac{1}{\sqrt{2}} S & 0 & P + \Delta 
    \end{matrix}
    \right) \,.
   \label{k.p Hamiltonian}
\end{align}
Here we have chosen the sign such that the hole effective masses are positive, and 
\begin{eqnarray}
   P &=& \frac{\hbar^2}{2m_0}\gamma_1 (k_x^2+k_y^2+k_z^2) \,, \\ 
   Q &=& \frac{\hbar^2}{2m_0}\gamma_2 (k_x^2+k_y^2-2k_z^2) \,, \\ 
   R &=& \frac{\hbar^2}{2m_0}\sqrt{3} [-\gamma_3(k_x^2-k_y^2)+2i\gamma_2 k_x k_y] \,, \\ 
   S &=& \frac{\hbar^2}{2m_0}2\sqrt{3} \gamma_3(k_x-ik_y)k_z \,. 
\end{eqnarray}
The above expressions of $P$, $Q$, $R$ and $S$ apply when the following correspondence holds between the reference and crystallographic axes \cite{Venitucci2018}: 
\begin{align}
 \hat{\boldsymbol{x}}   \parallel   [110], \quad \hat{\boldsymbol{ y}}  \parallel   [\bar 110], \quad \hat{\boldsymbol{ z}}   \parallel   [001] \,. 
\end{align} 
  
The L\"uttinger parameters $\lbrace \gamma_1, \gamma_2, \gamma_3 \rbrace$ are equal to $\lbrace 4.285, 0.339, 1.446 \rbrace$ for Si, and $ \lbrace 13.38, 4.24,  5.69 \rbrace$ for Ge; the spin--orbit parameter is $\Delta = 44$ meV for Si, and $\Delta = 290$ meV for Ge.

In the presence of an external electrostatic potential $V(\boldsymbol{r})$ that varies smoothly over the spatial scale of the lattice constant, the effective Hamiltonian for the low--energy hole states is given by the L\"uttinger--Kohn (LK) expression,
\begin{align}
H_{\rm LK} = H_{\boldsymbol{k} \cdot \boldsymbol{p}} + {\rm diag}\!\left[ V(\boldsymbol{r})\right] \,,
\label{LK Hamiltonian}
\end{align}
where the external potential is added to the diagonal elements of $H_{\boldsymbol{k} \cdot \boldsymbol{p}}$ [Eq.~\eqref{k.p Hamiltonian}]. The solution of the matrix Schr\"odinger equation determined by $H_{\rm LK}$ yields eigenvectors whose $\boldsymbol{r}$--dependent components are the envelope functions: we denote them as $\psi_{\alpha, b}(\boldsymbol{r})$, where $\alpha$ is the eigenstate index and $b \equiv (j,m)$ distinguishes the 6 components (bands). The total eigenstates, including the microscopic (Bloch) parts, are written as
\begin{align}
\big| \psi_{\alpha}  \big>   =  \int d \boldsymbol{r} \sum_b \psi_{\alpha, b}(\boldsymbol{r})  \big|  \varepsilon^+_{b}(\boldsymbol{r}) \big>    \,,
\label{single-particle eigenstates}
\end{align}
where
\begin{align}
\big|  \varepsilon^+_{b}(\boldsymbol{r}) \big> =   \sqrt{ \mathcal{V}_{\rm a}}    
\sum_{\boldsymbol{R}_k}     (-1)^k  \sum_{\xi, s_z} S_{b,\xi, s_z} 
\phi_{ p_{\xi}}(\boldsymbol{r} - \boldsymbol{R}_k) 
  \big|   \boldsymbol{r}, s_z \big> 
\label{Bloch states silicon}
\end{align}
is the Bloch state \cite{Voon2009, Secchi2020} for the band $b$. This combines $p$--type atomic orbitals $\phi_{ p_{\xi}}$ (with $\xi \in \lbrace x,y,z \rbrace$) centered on all the $N_{\rm a}$ atomic positions $\boldsymbol{R}_k$ of the crystal  (where $k \in \lbrace 0, 1 \rbrace$ labels the two atoms in each unit cell) with the spin states $s_z = \pm 1/2$, through the Clebsch--Gordan coefficients $S_{b,\xi,s_z}$; $\big| \boldsymbol{r}, s_z \big>$ is a position--spin eigenstate. The constant $\mathcal{V}_{\rm a}$ is the volume occupied by a single atom in the Si or Ge lattice, i.e., half of the two--atom unit cell; the normalization is chosen such that 
\begin{align}
\int d \boldsymbol{r} \sum_b \psi^*_{\alpha, b}(\boldsymbol{r}) \, \psi_{\alpha', b}(\boldsymbol{r})  = \delta_{\alpha, \alpha'} \,.
\end{align}

In this work, we focus on prototypical double quantum dots (DQDs), defined within a Si or Ge quantum well by means of electrostatic gates. The hole confinement is accounted for by the total potential  
\begin{align}
V(\boldsymbol{r}) = V_{\rm DQD}(x) + V_{\rm QD}(y) + V_{\parallel} \theta( |z| - L_z/2 ) \,,
\label{total QD potential}
\end{align}
where the last term accounts for the confinement in a well of width $L_z$ along the $z$ direction, with a height $V_\parallel$ that mimics the band offset between the semiconductor and the surrounding insulating materials.  
The confinement along the $x$ and $y$ directions is respectively given by the quartic and parabolic potentials
\begin{align}
& V_{\rm DQD}(x) = \frac{1}{2}\kappa   \frac{\left( x^2 - a^2 \right)^2}{4a^2}     \,,  \\
& V_{\rm QD}(y) = \frac{1}{2}\kappa  y^2  \,.
\end{align}
The minima of $V_{\rm DQD}(x)$ are located at $x = \pm a$, and are thus separated by a distance $D=2a$, to which we refer in the following as the {\it interdot distance}. The height of the interdot barrier is $V(\boldsymbol{0})=\kappa a^2 /8$. For $|x \mp a | \ll a$, the confinement along the $x$ direction is approximately harmonic, with the same spring constant as the one that characterizes the harmonic potential along the $y$ direction:
\begin{align}\label{eq01}
 V_{\rm DQD}(x) \approx \frac{1}{2} \kappa \left( x \mp a \right)^2  \,.
\end{align}

\subsubsection{Characterization of the single--hole states}

If the $\boldsymbol{r}$--dependence of the Bloch states in Eq.~\eqref{single-particle eigenstates} is neglected, the single--hole eigenstates can be rewritten as
\begin{equation}\label{eq: spin orbit}
    \big|\psi_\alpha \big> = \sum_{b} \big| \psi_{\alpha,b}\big> \otimes |b\rangle, 
\end{equation}
where the band states $|b\rangle$ replace the Bloch states and can be considered as spinors in this picture, and $\big< \boldsymbol{r} \big| \psi_{\alpha, b} \big> = \psi_{\alpha, b}(\boldsymbol{r})$. Hereinafter, we refer to envelope and band as the {\it orbital} and {\it spin} degrees of freedom, respectively.

The occupations of the six bands, whose sum is normalized to 1, are given by
\begin{equation}\label{eq:band occ}
    p_{\alpha,b} \equiv \langle \psi_{\alpha,b} |\psi_{\alpha,b} \rangle \,.
\end{equation}

\begin{figure}
    \centering
    \includegraphics[width=7cm]{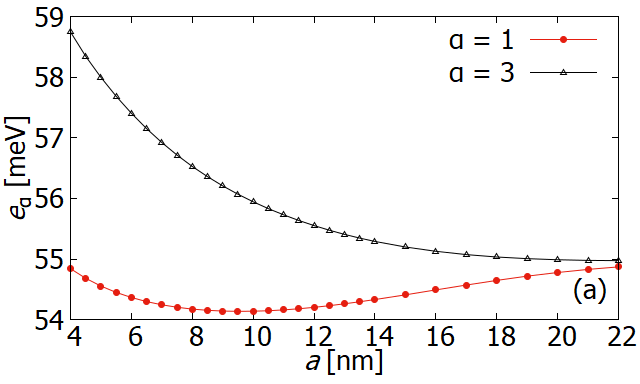}
    \vspace{0.2cm}
    \includegraphics[width=7cm]{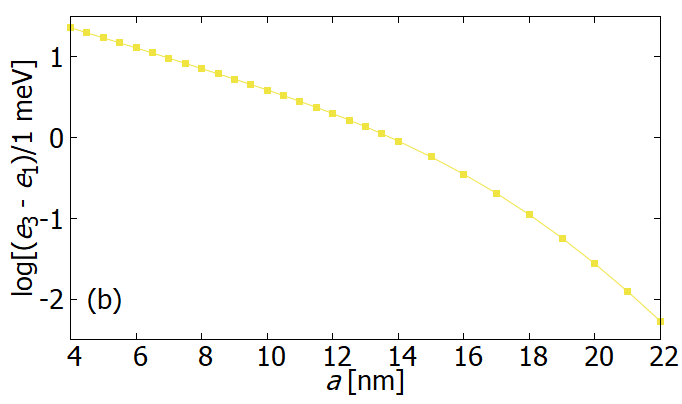}
    \vspace{0.2cm}
    \includegraphics[width=7cm]{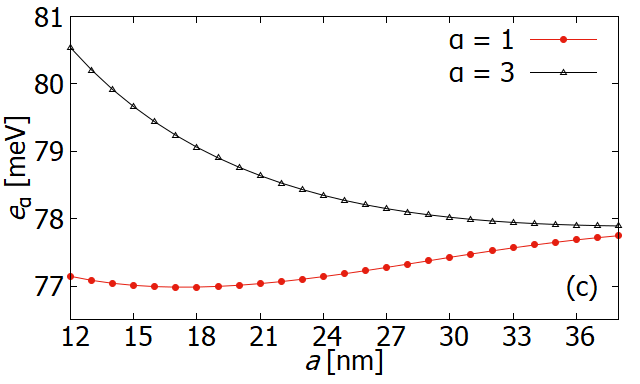}
    \vspace{0.2cm}
    \includegraphics[width=7cm]{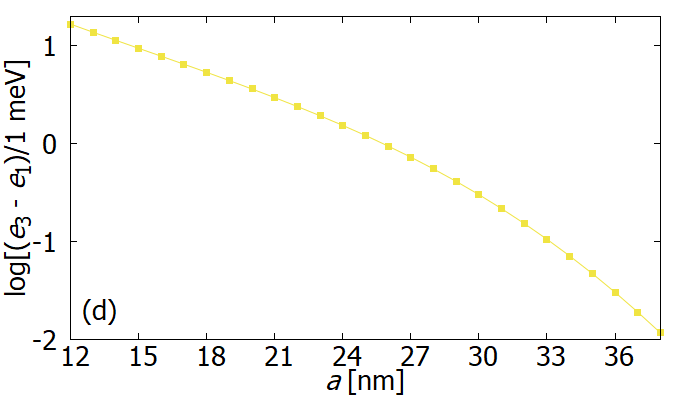}
    \caption{Energies $e_{\alpha}$ of the single--hole ground $(\alpha = 1)$ and first--excited $(\alpha=3)$ states in the (a) Si and (c) Ge double quantum dot, as functions of the half interdot distance $a=D/2$. Panels (b) and (d) report the differences $e_3-e_1$ for Si and Ge, respectively.}
    \label{fig:sp_energies}
\end{figure}

The eigenstates $|\psi_\alpha\rangle$ can be also characterized in terms of their spatial symmetries. In the following, we specifically refer to the expectation value of the operator $\sigma_{yz}$, which implements a reflection of the orbital states about the $yz$ plane, and is thus defined by the equation
\begin{equation}
\langle x,y,z | \sigma_{yz}|\psi_{\alpha}\rangle = \langle -x,y,z | \psi_{\alpha}\rangle .    
\end{equation}

\begin{figure}
    \centering
    \includegraphics[width=0.475\columnwidth]{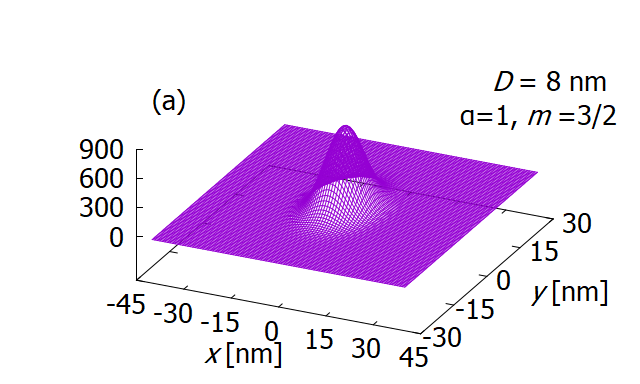}
    \includegraphics[width=0.475\columnwidth]{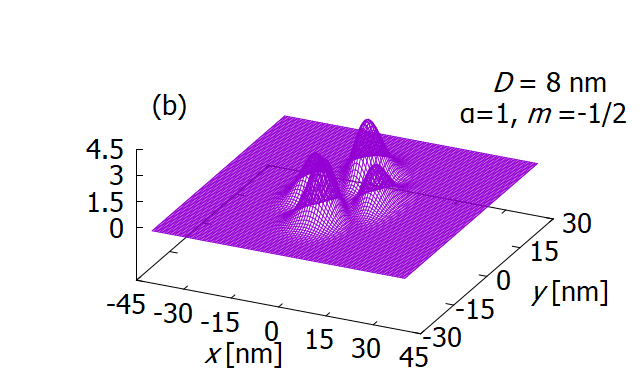}
    \includegraphics[width=0.475\columnwidth]{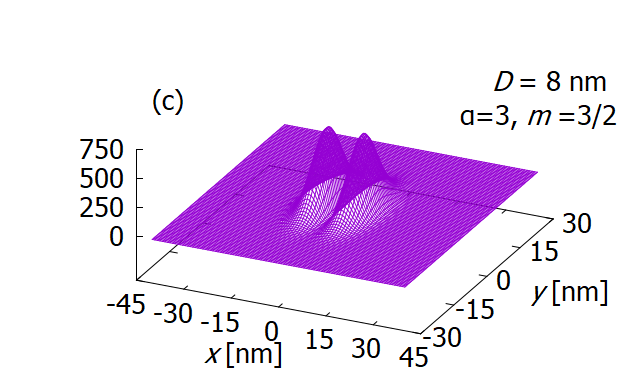}
    \includegraphics[width=0.475\columnwidth]{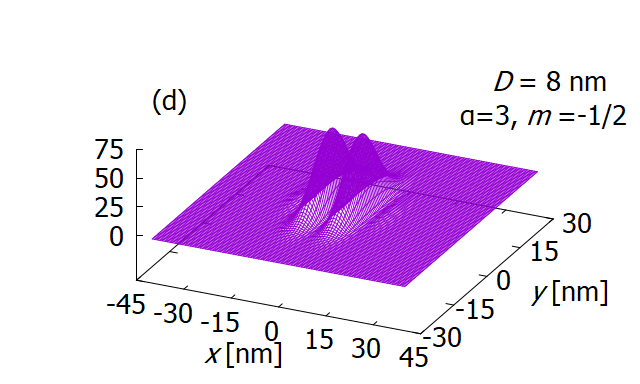}
    \includegraphics[width=0.475\columnwidth]{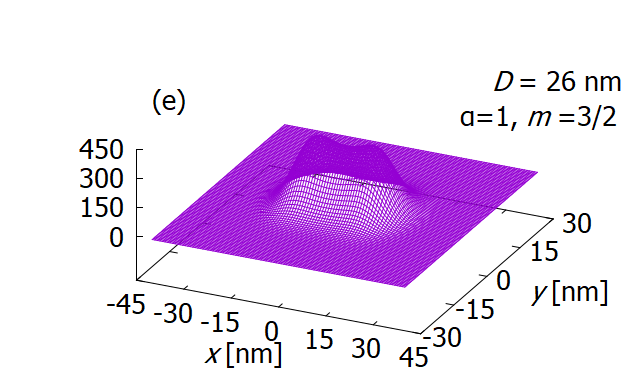}
    \includegraphics[width=0.475\columnwidth]{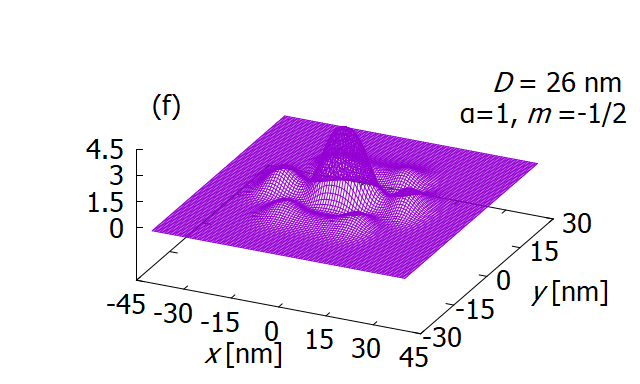}
    \includegraphics[width=0.475\columnwidth]{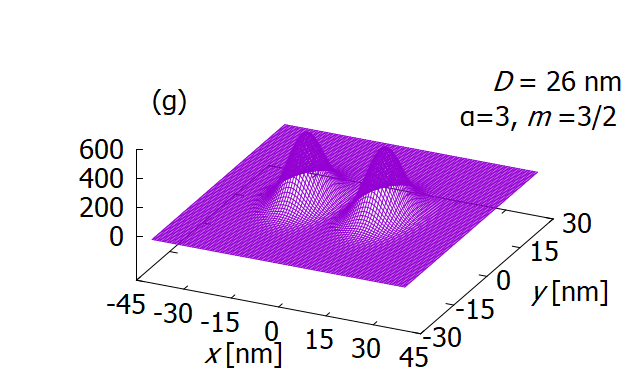}
    \includegraphics[width=0.475\columnwidth]{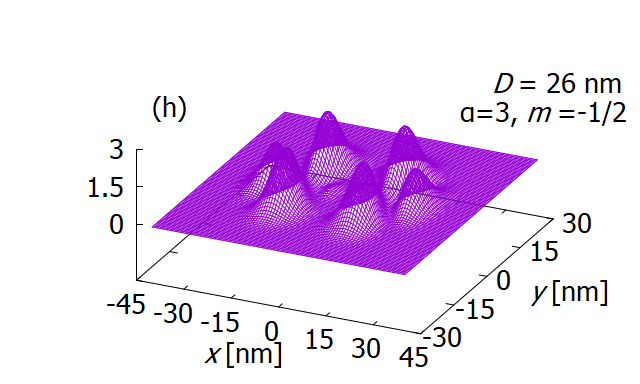}
    \caption{Profile along the in--plane directions, for $z=0$, of the band--resolved charge density $\rho^{\rm ch}_{\alpha, (3/2, m)} (\boldsymbol{r})$ in the Si dots (arbitrary units). Panels a-d and e-h refer to the representative interdot distances $D=8$ nm and $D=26$ nm, respectively. a, e: $(\alpha, m) = (1, 3/2)$; b, f: $(\alpha, m) = (1, -1/2)$; c, g: $(\alpha, m) = (3, 3/2)$; d, h: $(\alpha, m) = (3, -1/2)$. Since the states with $\alpha=2$ and $\alpha=4$ are the time--reversal conjugates of those with $\alpha=1$ and $\alpha=3$, respectively, one has $\rho^{\rm ch}_{2, (j, m)} = \rho^{\rm ch}_{1, (j, -m)}$ and $\rho^{\rm ch}_{4, (j, m)} = \rho^{\rm ch}_{3, (j, -m)}$.}
    \label{fig:molec_orbitals}
\end{figure}

In the single--hole states $|\psi_\alpha\rangle$, orbital and spin degrees of freedom are in general entangled. This means that it is {\it not} possible to write such states exactly in the factorized form
\begin{align}
| \psi^{\rm fact}_{\alpha} \rangle \equiv  |\psi_{\alpha,1}\rangle \otimes \sum_{b} c_{\alpha, b}    |b\rangle \,.
\label{factorized state}
\end{align}
%
The state in Eq.~\eqref{factorized state} is the product of an orbital function $| \psi_{\alpha, 1} \rangle$, common for all bands, and a spin state $\sum_{b} c_{\alpha, b} |b\rangle$, which is in general a linear combination of different band states. 

In general, the hole eigenstates display band mixing, i.e., different bands are coupled to different orbital functions. In order to investigate such mixing, we first compute the reduced single--spin density matrix:
\begin{equation}
    \rho^{{\rm sp}}_{\alpha} = \sum_{b, b'} \langle \psi_{\alpha,b'} |\psi_{\alpha,b} \rangle |b\rangle\langle b'| \,.
\end{equation}
Then, the entanglement between the orbital (envelope) and spin (band) degrees of freedom is quantified through the linear entropy of the reduced spin state:
\begin{equation}\label{eq:lin entr}
    S_L(\rho^{{\rm sp}}_{\alpha}) = 1 - \text{tr} \left[(\rho^{{\rm sp}}_{\alpha})^2 \right] = 1-\sum_{b,b'} |\langle \psi_{\alpha,b'} |\psi_{\alpha,b} \rangle|^2 \,,
\end{equation}
which ranges from $0$ to $5/6$, for increasing spin--orbit correlations and mixing of the density matrix $\rho^{{\rm sp}}_{\alpha}$. One has $S_L(\rho^{{\rm sp}}_{\alpha}) = 0$ if and only if the hole state $| \psi_{\alpha} \rangle$ has the factorized form given in Eq.~\eqref{factorized state}.

In this system, spin--orbit entanglement [$S_L(\rho^{{\rm sp}}_{\alpha})  \neq 0$] requires the fulfillment of two conditions. The first one is the occurrence of band mixing, i.e., the distribution of population among different bands. The second condition is that the orbital states corresponding to such different bands are not parallel to each other, i.e., that there are no constants $c_{\alpha, b}$ such that $|  \psi_{\alpha} \rangle$ can be factorized as in Eq.~\eqref{factorized state}.  

In order to separately quantify the effect of these two contributions, we compare the linear entropy of $\rho^{{\rm sp}}_{\alpha}$ with that of the fully--dephased reduced density operator
\begin{equation}
    \sigma^{{\rm sp}}_{\alpha} = \sum_{b} \langle \psi_{\alpha,b} |\psi_{\alpha,b} \rangle |b\rangle\langle b| \,,
\end{equation}
which is given by
\begin{equation}\label{eq:dep lin entr}
    S_L(\sigma^{{\rm sp}}_{\alpha}) = 1-\sum_{b} |\langle \psi_{\alpha,b} |\psi_{\alpha,b} \rangle|^2 \ge S_L(\rho^{{\rm sp}}_{\alpha}) \,.
\end{equation}
The linear entropy $S_L(\sigma^{{\rm sp}}_{\alpha})$ quantifies the band mixing alone, while the difference 
\begin{align}
S_L(\sigma^{{\rm sp}}_{\alpha}) -  S_L(\rho^{{\rm sp}}_{\alpha}) =      
    \sum_{b} \sum_{b' \neq b} \left| \langle \psi_{\alpha,b'} |\psi_{\alpha,b} \rangle \right|^2
\end{align} 
singles out the contribution to spin--orbit entanglement resulting from the different spatial dependencies of the orbital states corresponding to different bands. Low values of $S_L(\rho^{{\rm sp}}_{\alpha})$ imply that the state $| \psi_{\alpha} \rangle$ can be associated with a well--defined (non--mixed) spin state, either because the hole eigenstate presents a low degree of band mixing [relatively low values of $S_L(\sigma^{{\rm sp}}_{\alpha})$], or because the orbital states corresponding to different spin components are strongly overlapping [relatively high values of $S_L(\sigma^{{\rm sp}}_{\alpha})$].

\subsection{Numerical results}
\label{subsec:shs}

In the present Subsection, we report the properties of single--hole states in two Si and Ge horizontally coupled quantum dots, as a function of the interdot distance. In particular, single--hole energies are used to identify the parameter range corresponding to a weak interdot coupling, where the excitation energy associated with the motion along $x$ is smaller than that associated with the motion along $y$ and $z$, and the approximations underlying the Hubbard model apply. The comparison between the symmetry properties and spatial distributions of the different band components within each hole eigenstate provides a first representation of the correlation between spin and orbital degrees of freedom, which is quantitatively characterized by the linear entropies. Such characterization is propaedeutic to the introduction of the effective spin representation of single-- and two--hole states. 

\subsubsection{Interdot tunneling}

Within the present model, interdot tunneling is tuned by varying the parameter $a$ in the potential $V_{{\rm DQD}}(x,y)$. In fact, $a$ determines the distance $D=2a$ between the two potential minima ($x=\pm a$, $y=0$) that define the positions of the dots, as well as the height of the interdot barrier, given by $V_{{\rm DQD}}(0,0)=\kappa a^2 /8$. The strength of confinement is determined by $\kappa = m_0\omega^2/\gamma_1$, which can in turn be derived from the ``effective mass" $m_0/\gamma_1$ and from $\omega$. Here we set $\hbar\omega=5\,$meV for both Si and Ge dots, while the parameter $\gamma_1$, and thus the effective mass, is different in the two materials. Such difference can be quantified by the ratio between the characteristic length scales $l_\text{Ge}/l_\text{Si}\approx 1.767$, where $l \equiv (\hbar\gamma_1/m_0\omega)^{1/2}$. The parameter $\omega$ determines not only the interdot tunneling, but also the strength of the parabolic confinement along the $y$ direction. Finally, the confinement along the $z$ direction is induced by a potential well whose depth and width are given by $V_{\parallel}=4.0$ eV and $L_z=5\,$nm, respectively.

We are specifically interested in the regime of weak interdot coupling, where the energy scale associated with interdot tunneling is smaller than that associated with intradot excitations ($\delta$). In order to identify such regime, we compute the lowest energy eigenvalues $e_{\alpha}$ as a function of $a$ for both Si and Ge DQDs (Fig.~\ref{fig:sp_energies}). In the absence of an applied magnetic field, each energy level is doubly degenerate (Kramers degeneracy). The energy difference between the ground ($\alpha = 1,2$) and the first excited doublet ($\alpha = 3,4$) decreases faster than exponentially for increasing $a$, and drops below the intradot gap ($\delta_{\text{Si}}\approx 3.94\,$meV) for $a \approx 4\,$nm in the case of Si. In the case of Ge ($\delta_{\text{Ge}}\approx 4.58\,$meV), all considered values of $a$ correspond to a regime of weak interdot coupling.

In order to visualize the effect of interdot tunneling and the degree of orthogonality between the orbitals corresponding to different bands, we plot the profile along the $xy$ plane of the band--resolved charge density
\begin{align}
\rho^{\rm ch}_{\alpha, b} (\boldsymbol{r}) =  |\langle {\boldsymbol{r}} | \psi_{\alpha, b}\rangle |^2  \,,
\end{align}
for states $\alpha =1$ and $\alpha = 3$, and for two representative values of $D$ (Fig.~\ref{fig:molec_orbitals}). The overall character of each eigenstate is determined by the dominant heavy--hole contribution $(j,m)=(3/2,3/2)$, which is bonding and antibonding for the ground [panels (a, e)] and first excited (c, g) states, respectively. The main light--hole component $(j,m)=(3/2,-1/2)$ clearly displays a different spatial distribution, with a larger number of nodes (b, d, f, h). In particular, we observe a transition in the excited state: at small interdot distances, its heavy-- and light-- hole spatial distributions (c, d) are very similar, while at larger distances they are nearly disjointed (g, h). Such transition is captured in the dependence of the linear entropies on $a$ (see below). We omit the minority contributions to the charge density, which for states $\alpha = 1$ and $\alpha = 3$ correspond to $(j,m)=(3/2,-3/2)$, $(j,m)=(3/2,1/2)$ and $j = 1/2$, as they are comparatively negligible.

\begin{table}[]
\renewcommand{\arraystretch}{1.25}
    \centering
    \begin{tabular}{|l|c|c|c|c|}            
    \hline
      Si   & \multicolumn{2}{|c|}{$|\psi_1\rangle$} & \multicolumn{2}{|c|}{$|\psi_3\rangle$} \\
    \hline
      $a$ [nm]   & 4 & 22 & 4 & 22 \\
    \hline
    \hline
       $p_{\alpha,3/2, 3/2}$  &  0.989 & 0.904 &  0.895 & 0.919 \\
       $p_{\alpha,3/2,-1/2}$  &  0.006 & 0.0801 &  0.083 & 0.0643  \\
    \hline     
       $\langle \sigma_{yz} \rangle_\alpha$  &  0.996 & 1.00  &  $-0.973$ & $-1.00$  \\ 
    \hline
       $S_L(\rho_{\alpha}^{\text{sp}})$ &  0.0212 & 0.0216  & 0.0480 & 0.0207 \\
       $S_L(\sigma_{\alpha}^{\text{sp}})$ &  0.0217 & 0.0225  & 0.193 & 0.0207 \\
    \hline
    \end{tabular}
    \vspace{0.25cm}
    
    \caption{Characteristic quantities associated with single--hole states $|\psi_1\rangle$ and $|\psi_3\rangle$ of the Si DQD, for two representative values of $a$. The table displays: the two largest band occupations $p_{\alpha,b}$; 
    the expectation value of the symmetry operator $\sigma_{yz}$, i.e., $\langle \sigma_{yz}\rangle_\alpha = \langle\psi_\alpha |\sigma_{yz}|\psi_\alpha\rangle$; the linear entropy of the spin reduced density operators $\rho_{\alpha}^{{\rm sp}}$ and $\sigma_{\alpha}^{{\rm sp}}$. }
    \label{tab:single_hole}
\end{table}

\subsubsection{Band composition, symmetries and spin--orbit correlation}

Single--hole states can be characterized in terms of their band composition, the symmetry of their band components, and spin--orbit correlation.

In the case of Si, ground states $|\psi_1\rangle$ and $|\psi_2\rangle$ have a predominant heavy--hole character. At zero magnetic field, the individual eigenstates within each Kramers doublet are not unambiguously defined. In this set of calculations we break the Kramers degeneracy and unambiguously define the eigenstates by including a magnetic field of $10^{-2}\,$T along $z$, which is weak enough to avoid mixing between states belonging to different doublets (the hole coupling to the magnetic field is accounted for by means of Peierls's substitution and by including the Zeeman--Bloch Hamiltonian: see Ref.~\onlinecite{Bellentani2021} for details). Therefore, state $|\psi_1\rangle$ is characterized by a dominant $(j,m)=(3/2,3/2)$ component (while the main contribution to $|\psi_2\rangle$ corresponds to $m=-3/2$). The weight of such band decreases monotonically while the interdot distance increases, while that of the second largest contribution, namely $(j,m)=(3/2,-1/2)$, increases with increasing $D$; the opposite behavior characterizes state $|\psi_3\rangle$, belonging to the first excited Kramers doublet. In Table \ref{tab:single_hole} we report the values of the band occupations corresponding to the smallest and largest values of $a$ in the considered range. The occupation probabilities of the $m=-3/2$, $(j,m) = (3/2,1/2)$, and $j=1/2$ subbands (not reported) for states $|\psi_1\rangle$ and $|\psi_3\rangle$ are negligible. States $|\psi_2\rangle$ and $|\psi_4\rangle$ are the time--reversal conjugates of $|\psi_1\rangle$ and $|\psi_3\rangle$, respectively: therefore, $p_{2,j,m}=p_{1,j,-m}$ and $p_{4,j,m}=p_{3,j,-m}$.

Different orientations of the small magnetic field yield different occupations of the individual bands: for example, for $\boldsymbol{B}$ along the $x$ direction one has $p_{1,j,m}=p_{1,j,-m}$. This is due to the fact that the magnetic field, in the weak--field regime that we are considering, selects different linear combinations of the degenerate Kramers states, depending on its orientation. However, the total occupation of the heavy-- and light--hole bands for each Kramers doublet is independent of the field orientation.

\begin{figure}
    \centering
    \includegraphics[width=7cm]{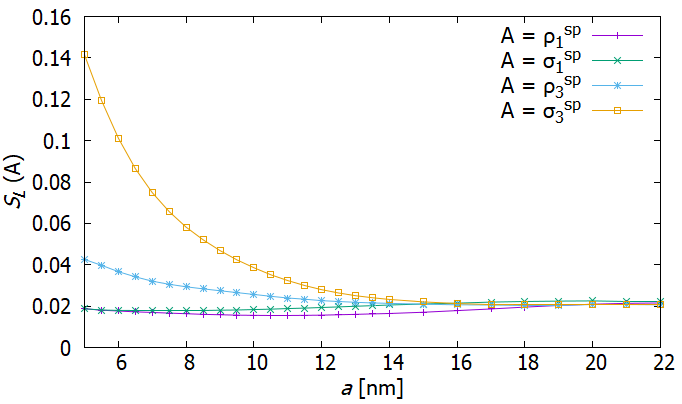}
    \caption{Linear entropies of the reduced spin density matrices $\rho^{{\rm sp}}_\alpha$ and their dephased counterparts $\sigma^{{\rm sp}}_\alpha$ for the hole eigenstates $|\psi_\alpha\rangle$ ($\alpha = 1,3$) of the Si DQD, as functions of the half interdot distance $a$. The maximum possible value for $S_L(\rho^{{\rm sp}}_\alpha)$ is $5/6 \approx 0.833$ .}
    \label{fig:lin entr}
\end{figure}

The expectation values of the mirror symmetry operator $\sigma_{yz}$ show that, in the considered range of parameter values, the ground and first excited molecular spin--orbitals can be labelled as spatially symmetric and antisymmetric states to a good degree of approximation. The band--resolved expectation values show that the symmetry along the $x$ direction is well defined and coincides for the subbands $(j,m)$ having the largest occupations, namely $b=(3/2,3/2)$ and $b=(3/2,-1/2)$ (while it is undefined for the remaining ones). In fact, we find that in these cases $ |\langle \psi_{\alpha,b} | \sigma_{yz} | \psi_{\alpha,b}\rangle| > 0.99\, \langle \psi_{\alpha,b} | \psi_{\alpha,b}\rangle $ in the whole considered range of values of $a$. The same inequality applies to the mirror symmetries about the two other coordinate planes.

We finally analyze the degree of entanglement between the spin and orbital degrees of freedom. In the case of Si (Fig.~\ref{fig:lin entr}), the ground state displays a low degree of mixing between the subbands [low $S_L(\sigma_1^{{\rm sp}})$], which also implies a small degree of entanglement between spin and orbital components [low $S_L(\rho_1^{{\rm sp}})$]. In the case of the first excited state, the difference between the two entropies is significant, especially for the smallest values of the interdot distance. This corresponds to a significant amount of band mixing, where, however, the orbital components corresponding to the two main subbands, $b=(3/2,3/2)$ and $b'=(3/2,-1/2)$, have a large overlap. In fact, the normalized overlap $ | \langle \psi_{3,b} | \psi_{3,b'}\rangle| (p_{3,b}\, p_{3,b'})^{-1/2} $ decreases monotonically from $0.88$ for $a=5\,$nm to less than $0.01$ for $a=22\,$nm. In any case, as the interdot distance varies, we either find a small amount of band mixing, or a large overlap between the orbitals corresponding to the main spin components. Therefore, a well--defined spin state can be assigned to each of the hole eigenstates.

In the case of Ge, the eigenstates $|\psi_\alpha\rangle$ (with $\alpha=1,2,3,4$) display symmetry properties similar to those of Si. However, the degree of band mixing is very limited, as the weight of the heavy--hole bands is higher than $0.999$ and $0.997$ respectively for the ground ($\alpha=1,2$) and first excited states ($\alpha=3,4$) in the whole investigated range of interdot distance values. Therefore, the linear entropies of both $\rho_\alpha^{{\rm sp}}$ and $\sigma_\alpha^{{\rm sp}}$ deviate negligibly from 0.

This degree of band mixing is not expected to change in the presence of strain, that can affect the Ge well in Si$_x$Ge$_{1-x}$ barriers. In particular, the uniaxial strain generated by the lattice mismatch between Ge and the Si$_x$Ge$_{1-x}$ substrate tends to increase the energy splittings between heavy--hole, light--hole and split--off bands, while it does not contribute to the off--diagonal (i.e., band--mixing) terms of the LK Hamiltonian (since $R_\epsilon=S_\epsilon=0$ for uniaxial strain, where $R_\epsilon$ and $S_\epsilon$ are the strain--dependent corrections to the terms $R$ and $S$, respectively, of the $\boldsymbol{k} \cdot \boldsymbol{p}$ Hamiltonian within Bir--Pikus's theory of strain). \cite{Terrazos2021}. Altogether, this results in a further suppression of the band mixing, and therefore does not qualitatively modify the above picture.

\section{Two--hole states}
\label{Sec: Two}

Our $6$--band calculations show that, for the considered confinement potentials, the occupation probability of the split--off bands ($j=1/2$) in the lowest--energy single--hole states is negligible. However, the split--off bands indirectly affect the overall band mixing, for they provide a coupling path between light-- and heavy--hole bands. Therefore, in order to simplify the following discussion, we project the single--hole states onto the $j=3/2$ subspace before proceeding with the calculation of the two--hole states. As a consequence, all hole states considered in the following have $j = 3/2$, and their individual spin state will be specified by the quantum number $m$ only.

\subsection{Method} 

\subsubsection{Diagonalization of the two--hole Hamiltonian}

In the case of two spin--$1/2$ fermions (such as electrons in one band with $l = 0$), one can combine the Slater determinants corresponding to different spin--orbitals to obtain a basis set formed by states with defined values of the total spin $J=S \in \lbrace 0, 1 \rbrace$. In the absence of spin--orbit interaction, each two--fermion state is the product either of a symmetric orbital and an antisymmetric spinor ($J = 0$), or of an antisymmetric orbital and a symmetric spinor ($J = 1$). The eigenstates of the Hamiltonian retain these symmetries.

In the case of spin--$3/2$ fermions, such as heavy and light holes in Si and Ge, it is still possible to combine the Slater determinants and obtain basis states that have a defined symmetry for both spin and orbital parts. In fact, one can introduce spinors $| J, M \rangle$, where $J \in \lbrace 0, 1, 2, 3 \rbrace$ corresponds to the total spin and $M=m_1+m_2$ is the eigenvalue of the two--hole spin--projection operator $J_z=j_{1,z}+j_{2,z}$. Spinors $| J, M \rangle$ are either symmetric or antisymmetric with respect to particle exchange, depending on whether $J$ is odd or even. Therefore, an odd--$J$ spinor must be combined with an antisymmetric orbital, and an even--$J$ spinor must be combined with a symmetric orbital, so that the product is antisymmetric. As a further complication with respect to the two--electron case, the hole Hamiltonian induces a mixing between symmetric and antisymmetric spinors, such that neither $J$ nor its parity are good quantum numbers (see below).

The basis set for the two--hole states is constructed from a set of orthonormal single--hole states $|\psi_\alpha\rangle$ [Eq.~\ref{eq: spin orbit}]. 
From these, one can define the two--hole Slater determinants 
\begin{eqnarray}
    |\Phi_{\alpha\beta} \rangle &=& 
\frac{1}{\sqrt{2}} \left( |\psi_\alpha \rangle |\psi_\beta \rangle-|\psi_\beta \rangle |\psi_\alpha \rangle \right) \nonumber\\
    &=& \sum_{\zeta = \pm 1} \sum_{m_1\ge m_2} |\Xi^{\alpha\beta\zeta}_{m_1,m_2} \rangle | \Upsilon^\zeta_{m_1,m_2} \rangle \,,
    \label{Slater det}
\end{eqnarray}
where
\begin{eqnarray}
    |\Upsilon^\zeta_{m_1 > m_2}\rangle&\equiv&\frac{1}{\sqrt{2}} \left( |m_1\rangle |m_2\rangle -\zeta|m_2\rangle |m_1 \rangle \right) \,, \\
    |\Upsilon^\zeta_{m_1 = m_2}\rangle&\equiv&\frac{1}{2}(1-\zeta) |m_1\rangle |m_2\rangle 
\end{eqnarray}
are the symmetric ($\zeta=-1$) and antisymmetric ($\zeta=+1$) combinations of the spinors $|m_1\rangle |m_2\rangle$ and $|m_2\rangle |m_1\rangle$. In the $j=1/2$ case (electrons), such combinations coincide respectively with the triplet and singlet states. In the present $j=3/2$ case, states $|\Upsilon^{-1}_{m_1 > m_2} \rangle$ and $|\Upsilon^{1}_{m_1 > m_2} \rangle$ belong respectively to the subspaces $J \in \lbrace 1,3 \rbrace$ and $J \in \lbrace 0,2 \rbrace$, but each of them generally includes components from different $J$ multiplets with the same parity. 

In order to guarantee the overall antisymmetry of the two--hole state, the orbitals corresponding to the $J \in \lbrace 0,2 \rbrace$ and $J \in \lbrace 1,3 \rbrace$ subspaces must be symmetric and antisymmetric, respectively:
\begin{align}
    |\Xi^{\alpha\beta\zeta}_{m_1 > m_2} \rangle & \equiv  \frac{1}{2} \big[
    |\psi_{\alpha,m_1} \rangle |\psi_{\beta,m_2} \rangle +\zeta |\psi_{\beta,m_2} \rangle |\psi_{\alpha,m_1} \rangle  \nonumber\\
    & \quad -  \zeta |\psi_{\alpha,m_2} \rangle |\psi_{\beta,m_1} \rangle - |\psi_{\beta,m_1} \rangle |\psi_{\alpha,m_2} \rangle\big]  , \nonumber \\
     |\Xi^{\alpha\beta\zeta}_{m_1 = m_2} \rangle  & \equiv  \frac{1- \zeta}{ 2\sqrt{2}} \big[
    |\psi_{\alpha,m_1} \rangle |\psi_{\beta,m_2} \rangle - |\psi_{\beta,m_2} \rangle |\psi_{\alpha,m_1} \rangle \big] .
\end{align}
The normalized states $|\Upsilon^\zeta_{m_1,m_2}\rangle$ defined in spin space are mutually orthogonal. The orbitals $|\Xi^{\alpha\beta\zeta}_{m_1,m_2}\rangle$ defined in real space, instead, are in general not normalized and not mutually orthogonal. 

The two--hole eigenstates are obtained by means of a Configuration--Interaction approach. The first step is the diagonalization of the single--hole LK Hamiltonian, which gives the eigenstates $|\psi_\alpha \rangle$ [Eq.~\ref{eq: spin orbit}], and the corresponding energy eigenvalues $e_{\alpha}$. From these, one constructs the basis for two--hole states, formed by the Slater determinants $|\Phi_{\alpha \beta}\rangle$ [Eq.~\eqref{Slater det}]. In this basis, the total Hamiltonian is given by:
\begin{align}
\langle \Phi_{\alpha \beta} | H | \Phi_{\alpha' \beta'} \rangle & = ( e_{\alpha} + e_{\beta}) \left( \delta_{\alpha, \alpha'} \delta_{\beta, \beta'} - \delta_{\alpha, \beta'} \delta_{\beta, \alpha'} \right) \nonumber \\
& \quad + V_{\rm C}(\alpha \beta, \alpha' \beta') \,,
\label{Hamiltonian matrix 2 holes}
\end{align} 
where the Coulomb term reads as
\begin{align}
  V_{\rm C}(\alpha \beta, \alpha' \beta') & = \sum_{m_1,m_2} \int d \boldsymbol{r}_1 d \boldsymbol{r}_2 \, \psi^*_{\alpha, m_1}(\boldsymbol{r}_1) \, \psi^*_{\beta,m_2}(\boldsymbol{r}_2)  \nonumber \\
& \quad  \times  V_{\rm C}(\boldsymbol{r}_1 - \boldsymbol{r}_2) \Big[ \psi_{\alpha', m_1}(\boldsymbol{r}_1) \psi_{\beta', m_2}(\boldsymbol{r}_2) \nonumber \\
& \quad - \psi_{\beta', m_1}(\boldsymbol{r}_1) \psi_{\alpha', m_2}(\boldsymbol{r}_2) \Big] \,,
\label{intraband Coulomb}
\end{align}
and $\psi_{\alpha, m} (\boldsymbol{r})= \langle \boldsymbol{r} |\psi_{\alpha, m}\rangle $. We assume here a uniform screening, $V_{\rm C}(\boldsymbol{r}_1 - \boldsymbol{r}_2) = \left( \epsilon |\boldsymbol{r}_1 - \boldsymbol{r}_2| \right)^{-1}$, quantified by the dielectric constants $\epsilon_{\rm Si} = 11.68$ for Si and $\epsilon_{\rm Ge} = 16.2$ for Ge. As we have shown in Ref.~\onlinecite{Secchi2020}, the Coulomb interaction also includes short--ranged interband processes, which can play a role in strongly confined systems. However, Eq.~\eqref{intraband Coulomb} is correct under the widely used approximation that the Coulomb potential is intraband. 

The diagonalization of this matrix yields the two--hole eigenstates as linear combinations of the basis Slater determinants,
\begin{align}
    |\Psi_k \rangle & \equiv \sum_{\alpha,\beta} C^k_{\alpha\beta} |\Phi_{\alpha\beta} \rangle
    \equiv \sum_{\zeta = \pm 1} \sum_{m_1\ge m_2} |\Pi_{m_1,m_2}^{k\zeta} \rangle  | \Upsilon^\zeta_{m_1,m_2} \rangle  \,,
\end{align}
where
\begin{align}
|\Pi_{m_1,m_2}^{k\zeta} \rangle \equiv \sum_{\alpha,\beta} C^k_{\alpha\beta} |\Xi^{\alpha\beta\zeta}_{m_1,m_2} \rangle \,,
\end{align}
and the coefficients $C^k_{\alpha \beta}$, which define the $k$--th eigenstate, result from the diagonalization of the two--hole Hamiltonian.

\subsubsection{Characterization of the two--hole states}

Analogously to the single--hole case, we define the spin reduced density matrix for the two--hole state $k$:
\begin{align}
\rho^{{\rm tp}}_{k} \equiv \sum_{\zeta = \pm 1} \sum_{\substack{m_1 \ge m_2\\ m_1'\ge m_2'}} \langle \Pi_{m_1',m_2'}^{k\zeta} | \Pi_{m_1,m_2}^{k\zeta} \rangle | \Upsilon^\zeta_{m_1,m_2} \rangle \langle \Upsilon^\zeta_{m_1',m_2'} |\,.
\end{align}
From the reduced density matrix one can derive the weights of the $J$ multiplets, which are given by:
\begin{align}\label{eq:J multiplets}
    p_{k,J} = \sum_{M=-J}^J p_{k,J,M} = \sum_{M=-J}^J \langle J,M| \rho^{{\rm tp}}_{k} | J,M \rangle \,. 
\end{align}
Since the length of the constituent spins is $j=3/2$, the possible values of $J$ range from $0$ to $3$.

The entanglement between spin and orbital components of the $k$--th two--hole eigenstate is quantified by the linear entropy
\begin{align}\label{eq:lin entr 2}
    S_L (\rho^{{\rm tp}}_{k}) = 1 - \text{tr} [(\rho^{{\rm tp}}_{k})^2] \,.
\end{align}
Analogously to the case of single--hole states, here spin--orbit entanglement requires the distribution of the reduced spin states among different two--hole spinors $(J,M)$. However, for a given occupation of these spinors, the amount of entanglement (and thus the linear entropy of $\rho^{{\rm tp}}_{k}$) depends on the overlap between the corresponding orbital states. In order to single out the contributions of these factors, we compare the linear entropy of the reduced spin state $\rho^{{\rm tp}}_{k}$ with that of its dephased counterpart
\begin{align}
 \sigma^{{\rm tp}}_{k} = \sum_{\zeta = \pm 1} \sum_{m_1\ge m_2 } \langle \Pi_{m_1,m_2}^{k\zeta} | \Pi_{m_1,m_2}^{k\zeta} \rangle | \Upsilon^\zeta_{m_1,m_2} \rangle \langle \Upsilon^\zeta_{m_1,m_2} | \,.
\end{align}
The dephasing is performed in the $(m_1,m_2)$ basis because the weights of the $| \Upsilon^\zeta_{m_1,m_2} \rangle \langle \Upsilon^\zeta_{m_1,m_2} |$ projectors differ from one another, if the orbitals corresponding to different subbands are not parallel. One has that 
\begin{equation}
    S_L (\sigma^{{\rm tp}}_{k}) \ge S_L (\rho^{{\rm tp}}_{k}),
\end{equation}
where the first entropy quantifies the mixing between different $(m_1,m_2)$ components, i.e., the spin--orbit entanglement that would be present if the orbitals corresponding to different bands were mutually orthogonal. The comparison between the above entropies thus allows to estimate the degree of orthogonality between the orbitals corresponding to different values of $(m_1,m_2)$.

\subsection{Numerical results}
\label{subsec: twoholenum}

In the present Subsection, we report the properties of the two--hole states in coupled quantum dots, as a function of the interdot distance. As in the case of single--hole states, the energies are used to identify the parameter range corresponding to a small yet finite splitting between the ground state (singlet) and the first excited multiplet (triplet). The reduced spin states, obtained after averaging over the envelope functions, represent the reference for validating the Hubbard model and the effective spin representation presented in the following sections. 

\begin{figure}
    \centering
    \includegraphics[width=7cm]{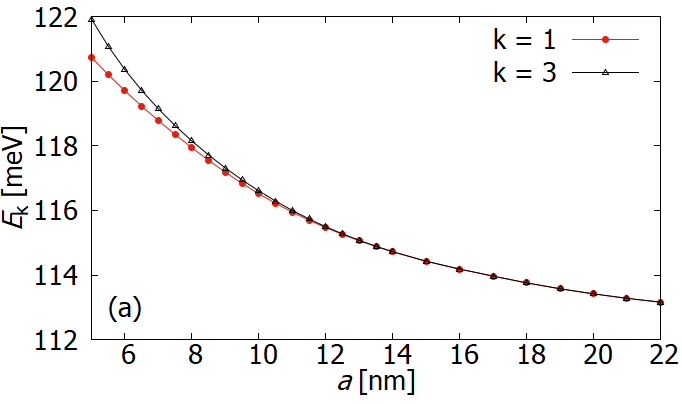}
    \includegraphics[width=7cm]{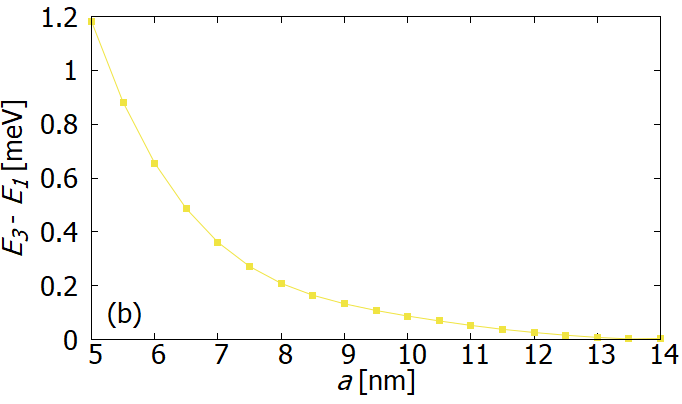}
    \includegraphics[width=7cm]{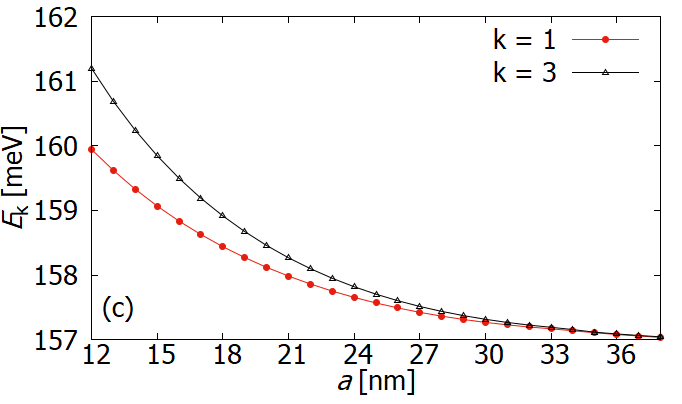}
    \includegraphics[width=7cm]{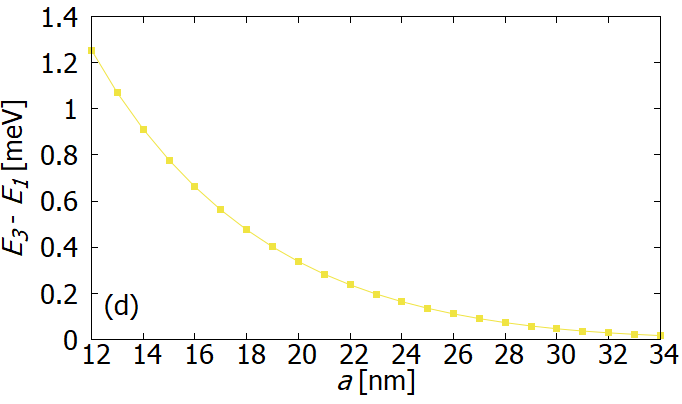}
    \caption{Energies of the lowest singlet ($E_1$) and of the three degenerate first--excited states ($E_2 = E_3 = E_4$), and singlet--triplet splitting ($E_3 - E_1$) as a function of the half interdot distance $a$ for the cases of Si (a, b) and Ge (c, d).}
    \label{fig:sing trip en}
\end{figure}

\begin{table}[]
\renewcommand{\arraystretch}{1.25}
    \centering
    \begin{tabular}{|l|c|c|c|c|}            
    \hline
      Si   & \multicolumn{2}{|c|}{$S\, (k=1)$} & \multicolumn{2}{|c|}{$T_0\,(k=3)$} \\
    \hline
      $a$ [nm]   & 5 & 14 & 5 & 14 \\
    \hline
    \hline
       $p_{k,0,0}$     &  0.490 & 0.487  &  0.000 & 0.000 \\
       $p_{k,2,0}$     &  0.477 & 0.488  &  0.000 & 0.006 \\
       $p_{k,2,\pm 2}$ &  0.012 & 0.001  &  0.008 & 0.012 \\
       \hline
       $p_{k,1,0}$     &  0.000 & 0.000  &  0.860 & 0.878 \\
       $p_{k,3,0}$     &  0.000 & 0.000  &  0.099 & 0.098 \\
       $p_{k,3,\pm 2}$ &  0.004 & 0.005  &  0.012 & 0.005 \\
    \hline
    \end{tabular}
    \vspace{0.25cm}
    
    \begin{tabular}{|l|c|c|c|c|}            
    \hline
      Ge   & \multicolumn{2}{|c|}{$S\, (k=1)$} & \multicolumn{2}{|c|}{$T_0\,(k=3)$} \\
    \hline
      $a$ [nm]   & 12 & 34 & 12 & 34 \\
    \hline
    \hline 
       $p_{k,0,0}$ & 0.499 & 0.499  &  0.000 & 0.000 \\
       $p_{k,2,0}$ &  0.499 & 0.499  &  0.000 & 0.000 \\
       $p_{k,2,\pm 2}$ &  0.000 & 0.000  &  0.000 & 0.000 \\
       \hline
       $p_{k,1,0}$ &  0.000 & 0.000 &  0.897 & 0.898 \\
       $p_{k,3,0}$ &  0.000 & 0.000 &  0.100 & 0.100 \\
       $p_{k,3,\pm 2}$ &  0.000 & 0.000 & 0.000 & 0.000 \\ 
    \hline
    \end{tabular}
    \caption{Main occupation probabilities $p_{k,J,M}$ corresponding to the $k=1$ (singlet $S$) and $k=3$ (triplet $T_0$) two--hole eigenstates and to the spin configuration $(J,M)$.}
    \label{tab:two_hole}
\end{table}

\subsubsection{Singlet and triplet}

The lowest--energy two--hole states at zero magnetic field consist of a ground singlet ($|S\rangle$, corresponding in the following to $k=1$) and an excited triplet (degenerate states $|T_- \rangle$, $|T_0\rangle$ and $|T_+\rangle$, corresponding respectively to $k=2$, $3$ and $4$). These degeneracies are analogous to those found in the single--band two--electron system. However, in that standard case singlet and triplet are eigenstates of $\boldsymbol{J}^2$ with eigenvalues $J=0$ and $J=1$, respectively, while the four--band two--hole system that we are treating is more complicated, as the eigenstates of the Hamiltonian are linear combinations of several two--particle spinors with different values of $J$. 

As the interdot distance increases, the singlet--triplet energy gap decreases faster than the tunneling--induced gap between the single--particle bonding and antibonding states, for both Si and Ge DQDs (Fig.~\ref{fig:sing trip en}). This is analogous to what is expected within a Hubbard--model picture, where the singlet--triplet energy splitting decays with increasing interdot distance as the second power of the hopping parameter.

\subsubsection{Reduced spin states}

We characterize the lowest two--hole eigenstates by means of the occupation probabilities $p_{k,J,M}$ of the spinors $(J,M)$ (Table \ref{tab:two_hole}). This analysis shows that the dominant part of the singlet state has $M=0$ and is given by comparable contributions from $J=0$ and $J=2$ spinors, which are antisymmetric with respect to permutation of the spin components $m_{1}$ and $m_{2}$. In the limit of vanishing band mixing, where the ground state has a purely heavy--hole character, the singlet state consists of a linear superposition of only these two contributions, with equal weights ($p_{1,0,0}=p_{1,2,0}=\tfrac{1}{2}$). Here, due to the presence of a small light--hole component in the single--particle states, the weights $p_{1,0,0}$ and $p_{1,2,0}$ slightly differ from one another, and minor contributions with $(J,M)=(2,\pm2)$ appear. Besides, contributions from odd--$J$ (symmetric) spinors appear, as is seen from $p_{1,3,\pm2} \neq 0$. Although the mixing between $J$ subspaces with different symmetries is here very limited, it is allowed by the symmetries of the Hamiltonian, and always expected to some degree if the single--hole states are not factorized (see the discussion of this topic in Sections \ref{Sec:Hubbard} and \ref{Sec: spin rep}). 

The triplet states are characterized by dominant contributions from $J=1$ and $J=3$ spinors, which have a symmetric character. In particular, the main terms of the $T_0$ state have $M=0$; in the single (heavy--hole) band limit, $p_{3,1,0} = \tfrac{9}{10}$ and $p_{3,3,0}=\tfrac{1}{10}$. The small light--hole component causes deviations from these limiting values, and the presence of additional contributions from the $(J,M)=(3,\pm2)$ states, as well as mixing with $J=2$ (antisymmetric) spinors. Finally, the $T_\pm$ states are characterized by odd values of $J$ and $M$, and tend to the maximally polarized states in the absence of a light--hole component ($p_{2,3,-3}=p_{4,3,3}=1$). 

\begin{figure}
    \centering 
     \includegraphics[width=8cm]{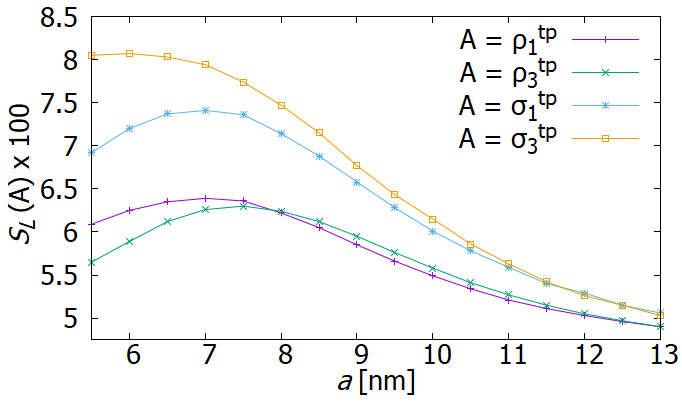}
    \caption{Linear entropies (multiplied by a factor $100$) of the reduced spin density matrices $\rho^{\rm tp}_{k}$ and of their dephased counterparts $\sigma^{\rm tp}_{k}$ for the singlet $(k = 1)$ and triplet $(k = 3)$ states in the Si double dot. For the three triplet states, $S_L(\rho^{\rm tp}_{2}) = S_L(\rho^{\rm tp}_{3}) = S_L(\rho^{\rm tp}_{4})$. The maximum possible value of $S_L(\rho^{\rm tp}_{k})$ is $15/16 =   93.75 \%$.}
    \label{fig:lin_entr}
\end{figure}

Overall, the dependence of the occupation probabilities $p_{1,J,M}$ and $p_{3,J,M}$ on the interdot distance is very weak, especially in the case of Ge. This is reflected in the dependence of the linear entropies on $a$ (see Fig. \ref{fig:lin_entr} for the case of Si). The linear entropy of the reduced two--spin density matrix $\rho^{\rm tp}_{k}$ is of the order of a few percents for both singlet and triplet. One can thus associate well--defined spin states to the four lowest eigenstates. The entropy would vanish and the limit of purely spin states would be achieved if the orbitals corresponding to the different spin states were parallel to each other. The opposite limit is the case where the orbitals corresponding to different bands are all mutually orthogonal. The reduced spin density matrix $\rho^{{\rm tp}}_{k}$ would then coincide with the dephased spin density matrix $\sigma^{{\rm tp}}_{k}$, whose linear entropy is also plotted in Fig.~\ref{fig:lin_entr} for a comparison. The difference between the linear entropy of $\sigma^{{\rm tp}}_{k}$ and that of $\rho^{{\rm tp}}_{k}$ thus represents an indicator of the orthogonality between the two--hole orbitals .

\section{Four--band Hubbard model}
\label{Sec:Hubbard}

We now build a semi--analytical Hubbard model that accounts for the main features of one-- and two--hole DQD states in Si and Ge. The model reproduces with a good degree of approximation the numerical results presented in the previous Section, and provides a more general and transparent physical picture, while including the complexity that originates from the four--band structure of the valence band. 

In the spirit of the Hubbard model \cite{Hubbard63}, the DQD single-- and two--hole states are derived from the single--hole eigenstates of the isolated dots. The two dots correspond to the regions close to the minima of the confinement potential $V(\boldsymbol{r})$. For sufficiently large values of the interdot distance $D=2 a$, the single--dot orbitals centered in different dots are approximately orthogonal and form a convenient basis for the description of the DQD states. The simplest model is obtained by considering only the ground Kramers doublet for each of the two dots, obtaining a basis of four single--hole states. The complication of the present model, with respect to standard (one--band) Hubbard models, lies in the fact that we take into account the four--band structure of the basis states.

\subsection{Derivation of the Hamiltonian}

As a first step, we consider the single--hole Hamiltonian and divide the DQD potential in three parts, in order to isolate the single--dot contributions:
\begin{align}
\hat{H}_{\rm DQD} = \hat{H}_{\boldsymbol{k} \cdot \boldsymbol{p}} + \underbrace{\hat{V}_1 + \hat{V}_2 + \hat{V}_{12}}_{\hat{V}_{\rm DQD}} \,.
\end{align}
Here $\hat{V}_{s}$, with $s \in \lbrace 1,2 \rbrace$, is the confining potential for dot $s$, and $\hat{V}_{12}  = \hat{V}_{\rm DQD} - \hat{V}_1 - \hat{V}_2$ is the tunneling--enabling potential. The single--dot terms $\hat{V}_{s}$ can be identified with the harmonic approximations of the DQD potential around its minima [Eq.~\ref{eq01}], i.e., $\hat{V}_{s} = \frac{1}{2} \kappa \left[ x - (-1)^s a \right]^2$. The definition of $\hat{V}_s$ is not univocal, but the formulation of the model is independent on its exact expression, as long as the symmetry between the two dots is preserved.

The single--hole states that form the ground Kramers doublet within each of the two dots satisfy the equation 
\begin{align}
\left( \hat{H}_{\boldsymbol{k} \cdot \boldsymbol{p}} + \hat{V}_s \right) \big| \psi_{s, \tau} \big> = E_{s, \tau} \big| \psi_{s, \tau} \big> \,,
\end{align}
and they can be written as
\begin{align}
&  \big| \psi_{ s, \Uparrow }  \big> = \sum_{m}    \big| \psi_{ s, \Uparrow,   m} \big> \otimes \big|  m \big>    \,,  \nonumber \\
&  \big| \psi_{ s, \Downarrow }  \big> = \sum_{m}   (-1)^{\frac{3}{2} -m} \big| \psi^*_{ s, \Uparrow,  -m} \big> \otimes \big|   m \big>    \,,  
\label{Down in terms of Up}
\end{align}
where $m \in \lbrace 3/2, 1/2, -1/2, -3/2 \rbrace$ is the eigenvalue of $j_z$, and the relation between $\big| \psi_{ s, \Uparrow }  \big>$ and $\big| \psi_{ s, \Downarrow }  \big>$ results from time--reversal symmetry. 
The index $\tau \in \lbrace \Uparrow, \Downarrow \rbrace $ specifies the eigenstate within the doublet and is akin to the eigenvalue of the third component of a pseudospin--$1/2$ operator, as we shall see. The excited states of each single dot are assumed to lie at energies high enough that they can be neglected in the development of a low--energy model for one-- and two--hole states. 

Numerical calculations performed on single--hole states of single dots suggest that, when $\boldsymbol{B} = (B_x, 0, 0)$ and $B_x \rightarrow 0$, the orbitals satisfy 
\begin{align}
\big|  \psi_{ s, \Uparrow,   m} \big> =  \big| \psi_{  s, \Uparrow,   -m} \big> \,, \quad
\big| \psi_{ s, \Downarrow,   m} \big> = - \big|  \psi_{ s, \Downarrow,  -m} \big> 
\label{eq03}
\end{align} 
to a very good approximation; this constraint conveniently reduces the size of the functional space. Besides, the imaginary parts of the orbitals are negligible with respect to the real parts. In the following, we develop and solve the model assuming the validity of Eq.~\eqref{eq03}, and then discuss the simplifications introduced by the further assumption that the orbitals are real (end of Section \ref{Sec: 1 hole Hubbard}, and Section \ref{sec: spinorbital wfcs}). These approximations will be justified {\it a posteriori}, by comparing the results derived from the Hubbard model with those obtained from the full numerical approach (see Appendix \ref{App: numerical 2 holes vs Hubbard}).
 
From Eq.~\eqref{eq03}, it follows that
\begin{align}
  \big| \psi_{ s, \Uparrow}  \big> &  \! \equiv \! \big| \psi_{ s, H} \big> \!\otimes\! \Big( \big|     \tfrac{3}{2} \big> \!+\! \big|   - \tfrac{3}{2} \big> \Big) 
   \!+\! \big| \psi_{ s, L} \big> \!\otimes\! \Big( \big|    \tfrac{1}{2} \big> + \big|  \! - \!\tfrac{1}{2} \big> \Big)   \,,  \nonumber \\
 \big| \psi_{ s, \Downarrow}  \big>  
  &  \!\equiv\!   \big| \psi^*_{ s, H} \big> \!\otimes\! \Big(   \big|   \tfrac{3}{2} \big>  \!-\!     \big|  - \tfrac{3}{2} \big> \Big)  
   \!-\! \big| \psi^*_{ s, L} \big> \!\otimes\! \Big(   \big|   \tfrac{1}{2} \big> \! -\!     \big|  - \tfrac{1}{2} \big> \Big)    \,,  
   \label{single-hole basis}
\end{align}
where $\big| \psi_{ s, H} \big> \equiv \big| \psi_{ s, \Uparrow,   \frac{3}{2}} \big> $ and $\big| \psi_{ s, L} \big> \equiv \big| \psi_{ s, \Uparrow,   \frac{1}{2}} \big> $ are the heavy-- and light--hole orbitals for dot $s$.

If the interdot distance is large compared to the extension along $x$ of the single--dot eigenstates $\big| \psi_{ s, \tau}  \big>$, one has that
\begin{align}
\hat{V}_{2} \big| \psi_{1, \tau} \big> \approx 0 \,, \quad   \hat{V}_{1} \big| \psi_{2, \tau} \big> \approx 0 \,, \quad  \big< \psi_{1, \tau} \big| \psi_{2, \tau'} \big> \approx 0 \,,
\end{align}
and the matrix elements of the single--hole Hamiltonian are
\begin{align}
\big< \psi_{s, \tau} \big| \hat{H}_{\rm DQD} \big| \psi_{s', \tau'} \big> = \delta_{\tau, \tau'} \left[\delta_{s,s'} E_{s, \tau} + T^{(\tau)}_{s, s'}\right] \,,
\end{align}
where
\begin{align}
 T^{(\Uparrow)}_{s, s'} & =  2 \sum_{h \in \lbrace H, L \rbrace} \int d \boldsymbol{r} V_{12}(\boldsymbol{r})         \psi_{ s, h}(\boldsymbol{r}) \,  \psi_{ s', h} (\boldsymbol{r}) \nonumber\\
& = \big< \psi_{s, \Uparrow} \big| \hat{V}_{12} \big| \psi_{s', \Uparrow} \big> = \big< \psi_{s,\Downarrow} \big| \hat{V}_{12} \big| \psi_{s',\Downarrow} \big>^* \,, \nonumber \\
 T^{(\Downarrow)}_{s, s'} & =  \left[ T^{(\Uparrow)}_{s, s'} \right]^* \,.
\end{align}

In second quantization, the single--hole Hamiltonian resulting from the above considerations reads as
\begin{align}
\hat{H}_{\rm DQD} = \sum_{\tau} \sum_{s,s'} \left[ \delta_{s,s'} E_{s, \tau} + T_{s,s'}^{(\tau)} \right] \hat{\psi}^{\dagger}_{s, \tau} \hat{\psi}_{s', \tau} \,.
\end{align}
The on--site single--hole parameters ($E_{1, \Uparrow} = E_{2, \Uparrow} = E_{1, \Downarrow} = E_{2, \Downarrow}$ and $T^{(\Uparrow)}_{1,1} = T^{(\Uparrow)}_{2,2} = T^{(\Downarrow)}_{1,1} = T^{(\Downarrow)}_{2,2}$) only add a constant contribution to the eigenvalues, and can thus be set to 0. The only relevant single--hole parameter is the one related to inter--site hopping:
\begin{align}
T^{(\Uparrow)}_{1,2} = \left[ T^{(\Uparrow)}_{2,1} \right]^* = \left[ T^{(\Downarrow)}_{1,2} \right]^* = T^{(\Downarrow)}_{2,1} \quad \equiv T \,.
\end{align}

In order to obtain an analytically solvable model, we assume that the interaction Hamiltonian is intraband and on--site. Its expression is thus given by 
\begin{align}
\hat{H}_U = \frac{1}{2} \sum_s \sum_{ \tau_1, \tau_2, \tau_3, \tau_4}   U_{\tau_1, \tau_2, \tau_3, \tau_4} \hat{\psi}^{\dagger}_{s, \tau_1} \hat{\psi}^{\dagger}_{s, \tau_2} \hat{\psi}_{s, \tau_3} \hat{\psi}_{s, \tau_4} \,,
\end{align}
where 
\begin{align}
U_{\tau_1, \tau_2, \tau_3, \tau_4}   & = \sum_{m,m'}\int d \boldsymbol{r} \int d \boldsymbol{r}' \, \psi^*_{s, \tau_1, m}(\boldsymbol{r})  \psi^*_{s, \tau_2, m'}(\boldsymbol{r}') \nonumber \\
& \quad \times V_{\rm Coulomb}(\boldsymbol{r} - \boldsymbol{r}') \,  \psi_{s, \tau_3, m'}(\boldsymbol{r}')  \psi_{s, \tau_4, m}(\boldsymbol{r}) \nonumber \\
& \equiv \delta_{\tau_1, \tau_4} \delta_{\tau_2, \tau_3} U \,,
\end{align} 
and
\begin{align}
U & \equiv 4 \int d \boldsymbol{r} \int d \boldsymbol{r}' \, \left\{ \left|  \psi_{ s, H}(\boldsymbol{r}) \right|^2 + \left| \psi_{ s, L}(\boldsymbol{r}) \right|^2    \right\} \nonumber \\
& \quad \times V_{\rm Coulomb}(\boldsymbol{r} - \boldsymbol{r}') \,  \left\{ \left|  \psi_{ s, H}(\boldsymbol{r}') \right|^2 + \left| \psi_{ s, L}(\boldsymbol{r}') \right|^2    \right\}  \,.
\end{align}

The two--site Hubbard Hamiltonian resulting from the previous considerations is
\begin{align}
\hat{H}_{\rm Hubbard} & = \left[ T  \left( \hat{\psi}^{\dagger}_{1, \Uparrow} \hat{\psi}_{2, \Uparrow} +
\hat{\psi}^{\dagger}_{2, \Downarrow} \hat{\psi}_{1, \Downarrow} \right)
+ \text{h.c.} \right]
  \nonumber \\
& \quad + U \left( \hat{n}_{1, \Uparrow} \hat{n}_{1, \Downarrow} + \hat{n}_{2, \Uparrow} \hat{n}_{2, \Downarrow}  \right)  \,,
\end{align}
where $\hat{n}_{s, \tau} = \hat{\psi}^{\dagger}_{s, \tau} \hat{\psi}_{s, \tau}$, and the hopping parameter $T= |T| {\rm e}^{{\rm i} \theta}$ is, in general, complex.

The two--site Hubbard model can be solved analytically at any occupation number $N \in \lbrace 0,1,2,3,4 \rbrace$. We are interested here in the cases of $N = 1$ and $N = 2$.

\subsection{One--hole solutions of the Hubbard model}
\label{Sec: 1 hole Hubbard}

In the case of $N = 1$, there are two energy eigenvalues $e_\pm = \pm |T|$, both doubly degenerate. The eigenstates 
\begin{align}
\big| \psi_{\pm, \tau} \big> & = \frac{1}{\sqrt{2}} \left(   \hat{\psi}^{\dagger}_{1, \tau}   \pm {\rm e}^{- {\rm i} \theta } \hat{\psi}^{\dagger}_{2, \tau}  \right) \big| 0 \big>    
\end{align}
are explicitly given by the expressions
\begin{align}
|\psi_{\pm, \Uparrow}\rangle 
& =  \frac{1}{\sqrt{2}} \Big[ \left( |\psi_{1, H}\rangle \pm {\rm e}^{- {\rm i} \theta }  | \psi_{2, H} \rangle \right)      \otimes \left( \left| \tfrac{3}{2} \right> + \left| -\tfrac{3}{2} \right> \right) \nonumber \\
& \quad 
 +  \left(   | \psi_{1, L}\rangle  \pm {\rm e}^{- {\rm i} \theta } | \psi_{2, L}\rangle \right)   \otimes \left( \left| \tfrac{1}{2} \right> +  \left| -\tfrac{1}{2} \right> \right)  \Big]  \,, \nonumber \\
|\psi_{\pm, \Downarrow}\rangle 
& =  \frac{1}{\sqrt{2}} \Big[  \left(     |\psi^{*}_{1, H}\rangle \pm {\rm e}^{- {\rm i} \theta }  | \psi^{*}_{2, H} \rangle \right)      \otimes \left( \left| \tfrac{3}{2} \right> -  \left| -\tfrac{3}{2} \right> \right) \nonumber \\
& \quad 
 - \left(   | \psi^{*}_{1, L}\rangle  \pm {\rm e}^{- {\rm i} \theta } | \psi^{*}_{2, L}\rangle \right)   \otimes \left(  \left| \tfrac{1}{2} \right> -  \left| -\tfrac{1}{2} \right>  \right)  \Big]  \,.
 \label{SP Hubbard states}
\end{align}

Under the (numerically justified) approximation that the single--dot orbitals are real, the hopping parameter $T$ is real as well. In this case, it follows from Eqs.~\eqref{SP Hubbard states} that, despite the complexity associated with the presence of four bands, the orbital wave functions contributing to each single--hole eigenstate of the Hubbard model have the same, well--defined molecular character (bonding or anti--bonding), analogously to what is found in single--band systems. This is consistent with our numerical results in the regime of weak coupling between the dots, as discussed in Subsec.~\ref{subsec:shs}. According to Eqs.~\eqref{SP Hubbard states}, the sign of $T$ (i.e., whether ${\rm e}^{- {\rm i} \theta } = -1$ or $+1$) determines the ordering of the single--hole eigenstates; our numerical results are consistent with $T < 0$ (i.e., ${\rm e}^{- {\rm i} \theta } = -1$), yielding a bonding ground state, as in natural diatomic molecules. We mention that in different $4$--band systems, such as vertically coupled DQDs in InGaAs/GaAs, also the alternative possibility of having $T > 0$ and an anti--bonding ground state has been reported \cite{Climente2008a, Doty2008, Yakimov2012}.

\subsection{Two--hole solutions of the Hubbard model}

In the case of $N = 2$, it is convenient to write the basis states as
\begin{align}\label{eq02}
& \big| \mathbb{S} \big > \equiv \frac{1}{\sqrt{2}} \left( \hat{\psi}^{\dagger}_{1, \Uparrow} \hat{\psi}^{\dagger}_{2, \Downarrow}
- \hat{\psi}^{\dagger}_{1, \Downarrow} \hat{\psi}^{\dagger}_{2, \Uparrow} \right) \big| 0 \big> \,, \nonumber \\
& \big| \mathbb{T}_{\Uparrow} \big > \equiv \hat{\psi}^{\dagger}_{1, \Uparrow} \hat{\psi}^{\dagger}_{2, \Uparrow} \big| 0 \big> \,, \nonumber \\
& \big| \mathbb{T}_0 \big > \equiv \frac{1}{\sqrt{2}} \left( \hat{\psi}^{\dagger}_{1, \Uparrow} \hat{\psi}^{\dagger}_{2, \Downarrow}
+ \hat{\psi}^{\dagger}_{1, \Downarrow} \hat{\psi}^{\dagger}_{2, \Uparrow} \right) \big| 0 \big> \,, \nonumber \\
& \big| \mathbb{T}_{\Downarrow} \big > \equiv \hat{\psi}^{\dagger}_{1, \Downarrow} \hat{\psi}^{\dagger}_{2, \Downarrow} \big| 0 \big> \,, \nonumber \\
& \big| \mathbb{D}_- \big> \equiv  \frac{1}{\sqrt{2}}   \left( \hat{\psi}^{\dagger}_{1, \Uparrow} \hat{\psi}^{\dagger}_{1, \Downarrow}  - \hat{\psi}^{\dagger}_{2, \Uparrow} \hat{\psi}^{\dagger}_{2, \Downarrow} \right) \big| 0 \big> \,, \nonumber \\
& \big| \mathbb{D}_+ \big> \equiv  \frac{1}{\sqrt{2}}   \left( \hat{\psi}^{\dagger}_{1, \Uparrow} \hat{\psi}^{\dagger}_{1, \Downarrow}  + \hat{\psi}^{\dagger}_{2, \Uparrow} \hat{\psi}^{\dagger}_{2, \Downarrow} \right) \big| 0 \big> \,.
\end{align}

From the diagonalization of the Hubbard Hamiltonian on this basis, it is found that the six eigenstates are distributed among four distinct energy levels (three singlets and one triplet). In terms of the exchange energy
\begin{align}
J \equiv  \frac{   \sqrt{U^2 + 16 \left| T \right|^2} -  U }{ 2 } \,,
\label{Hubbard J}
\end{align}
the lowest energy level is $E =  - J$ and it corresponds to the eigenstate
\begin{align}
\left|   - J \right>   = \frac{  2 \left| T \right| \Big[ - \cos(\theta)   \big| \mathbb{S} \big>  - {\rm i} \, \sin(\theta)   \big| \mathbb{T}_0 \big> \Big]  -  E_{ -} \big| \mathbb{D}_+ \big>  }{\sqrt{4 \left| T \right|^2 + \left( E_{-}   \right)^2 }}  \,.
\end{align}
The energy level $E = 0$ corresponds to the following three degenerate eigenstates:
\begin{align}
& \left| T_{x,0} \right> = - \cos(\theta) \big| \mathbb{T}_0 \big> - {\rm i} \, \sin(\theta)  \big| \mathbb{S} \big> \,, \nonumber\\
& \left| T_{x,+} \right> =\left| \mathbb{T}_{\Uparrow} \right> \,, \quad
 \left| T_{x,-}  \right> =  \left| \mathbb{T}_{\Downarrow} \right> \,.
\end{align}
The energy level $E = U$ corresponds to the eigenstate
\begin{align}
\left|   U \right> =  \big| \mathbb{D}_- \big>  \,,  
\end{align}
the eigenvalue $E =  U + J$ corresponds to the eigenstate
\begin{align}
\left|  U + J \right>   = \frac{  2 \left| T \right| \Big[ - \cos(\theta)   \big| \mathbb{S} \big>  - {\rm i} \, \sin(\theta)   \big| \mathbb{T}_0 \big> \Big]  -  E_{ +} \big| \mathbb{D}_+ \big>  }{\sqrt{4 \left| T \right|^2 + \left( E_{+}   \right)^2 }} \,.
\end{align}   
It should be noticed that, in the case of a real hopping parameter ($\theta \in \lbrace 0,\pi \rbrace$), the $\big| \mathbb{T}_0 \big>$ components in the singlets vanish, just like the $\big| \mathbb{S} \big>$ component in the triplet. One thus recovers the formal results of the standard Hubbard model.

\subsection{Spin--orbital wave functions for the low--energy two--hole eigenstates}
\label{sec: spinorbital wfcs}

In order to gain a deeper physical understanding of the two--hole eigenstates, we determine the corresponding spin--orbital wave functions. By combining Eqs.~\eqref{single-hole basis} and \eqref{eq02}, one can write explicit expressions for the two--hole eigenstates of the Hubbard model, displaying the four--band structure of the envelope functions. The resulting expressions are unwieldy in the most general case, but can be simplified by assuming that the orbital wave functions are real, as we already did in the case of $N=1$ (see the end of Section \ref{Sec: 1 hole Hubbard}). In the limit of $|T| \ll U$, one can neglect the double--occupancy contributions to the ground state, and
\begin{align}
\big| - J \big > \approx \big| \mathbb{S} \big> \,.
\end{align}
The lowest singlet and triplet states thus coincide with the (single--occupancy) eigenstates of an effective pseudospin--$1/2$ Heisenberg Hamiltonian,
\begin{align}
\hat{H}_{\rm Heisenberg} = J \left( \hat{\boldsymbol{S}}_1 \cdot \hat{\boldsymbol{S}}_2 - \frac{1}{4} \right) \,,
\end{align}
where
\begin{align}
\hat{\boldsymbol{S}}_s = \frac{1}{2}\sum_{\tau, \tau' \in \lbrace \Uparrow, \Downarrow \rbrace} \hat{\psi}^{\dagger}_{s,\tau} \boldsymbol{\sigma}_{\tau, \tau'} \hat{\psi}_{s, \tau'} \,,
\end{align}
is the pseudospin operator for site $s \in \lbrace 1,2 \rbrace$, and $\boldsymbol{\sigma}$ is the vector of Pauli matrices. 
 
While the ground state is uniquely determined, the triplet states are degenerate; therefore, given a solution for those three states, we can equivalently consider any set of three orthogonal linear combinations of them. To facilitate the comparison with numerical results (see Sections \ref{subsec:shs} and \ref{subsec: twoholenum}, and Appendix \ref{App: numerical 2 holes vs Hubbard}), we look for linear combinations of the triplet states that are compatible with those found numerically at $\boldsymbol{B} = (0,0,B_z)$, $B_z \rightarrow 0$. We find that, under the assumption that envelope functions are real, the sought--after combinations are
\begin{align}
& \big| T_{z,0} \big> \equiv  \frac{1}{\sqrt{2}}   \big| \mathbb{T}_{\Uparrow}  \big>   -  \frac{1}{\sqrt{2}} \big| \mathbb{T}_{\Downarrow} \big>   \,, \nonumber \\
& \big| T_{z,+} \big> \equiv \frac{1}{2} \big| \mathbb{T}_{\Uparrow}   \big> + \frac{1}{\sqrt{2}} \big| \mathbb{T}_0 \big>  + \frac{1}{2} \big| \mathbb{T}_{\Downarrow} \big>  \,, \nonumber \\
& \big| T_{z,-} \big> \equiv \frac{1}{2} \big| \mathbb{T}_{\Uparrow}  \big> - \frac{1}{\sqrt{2}} \big| \mathbb{T}_0 \big>  + \frac{1}{2} \big| \mathbb{T}_{\Downarrow} \big>  \,.
\end{align}
 
We switch from the spinor basis $|m_1, m_2\rangle$ to the spinor basis $|J,M\rangle$, where $M = m_1 + m_2$ and $J \in \lbrace 0,1,2,3 \rbrace$. The change of basis is achieved through the Clebsch--Gordan transformation, see Table \ref{tab:my_label}. We also introduce the condensed notation
\begin{align}
& |\Psi^S_{sh, s'h'}\rangle \equiv |\psi_{sh}\rangle |\psi_{s'h'}\rangle + |\psi_{s'h'}\rangle |\psi_{sh}\rangle\,, \nonumber \\ 
& |\Psi^A_{sh, s'h'}\rangle \equiv |\psi_{sh}\rangle |\psi_{s'h'}\rangle - |\psi_{sh}\rangle |\psi_{s'h'}\rangle\,,
\end{align}
which identifies symmetric and anti--symmetric two--hole orbitals; here $h, h' \in \lbrace H, L \rbrace$ denote the hole bands among those included in the model (heavy or light). 

\begin{table}[]
    \centering
    \begin{tabular}{c|cccc}
     $|J,M\rangle$    & $S^{J}_{3/2,M-3/2}$ & $S^{J}_{1/2,M-1/2}$ & $S^{J}_{-1/2,M+1/2}$ & $S^{J}_{-3/2,M+3/2}$ \\
    \hline
    $|0,0\rangle$     & $-1/2$ & $1/2$ & $-1/2$ & $1/2$ \\
    $|1,1\rangle$     & $\sqrt{3}/\sqrt{10}$ & $-2/\sqrt{10}$ & $\sqrt{3}/\sqrt{10}$ & $-$ \\
    $|1,0\rangle$     & $3/\sqrt{20}$ & $-1/\sqrt{20}$ & $-1/\sqrt{20}$ & $3/\sqrt{20}$ \\
    $|1,-1\rangle$    & $-$ & $-\sqrt{3}/\sqrt{10}$ & $2/\sqrt{10}$ & $-\sqrt{3}/\sqrt{10}$ \\
    $|2,2\rangle$     & $-1/\sqrt{2}$ & $1/\sqrt{2}$ & $-$ & $-$ \\
    $|2,1\rangle$     & $-1/\sqrt{2}$ & $-$ & $1/\sqrt{2}$ & $-$ \\
    $|2,0\rangle$     & $-1/2$ & $-1/2$ & $1/2$ & $1/2$ \\
    $|2,-1\rangle$     & $-$ & $1/\sqrt{2}$ & $-$ & $-1/\sqrt{2}$ \\
    $|2,-2\rangle$     & $-$ & $-$ & $-1/\sqrt{2}$ & $1/\sqrt{2}$ \\
    $|3,3\rangle$     & $1$ & $-$ & $-$ & $-$ \\
    $|3,2\rangle$     & $1/\sqrt{2}$ & $1/\sqrt{2}$ & $-$ & $-$ \\
    $|3,1\rangle$     & $-1/\sqrt{5}$ & $-\sqrt{3}/\sqrt{5}$ & $-1/\sqrt{5}$ & $-$ \\
    $|3,0\rangle$     & $1/\sqrt{20}$ & $3/\sqrt{20}$ & $3/\sqrt{20}$ & $1/\sqrt{20}$ \\
    $|3,-1\rangle$     & $-$ & $1/\sqrt{5}$ & $\sqrt{3}/\sqrt{5}$ & $1/\sqrt{5}$ \\
    $|3,-2\rangle$     & $-$ & $-$ & $1/\sqrt{2}$ & $1/\sqrt{2}$ \\
    $|3,-3\rangle$     & $-$ & $-$ & $-$ & $1$ \\
    \end{tabular}
    \caption{Clebsch--Gordan coefficients $ S^J_{m_1 , M - m_1} $ defining the decomposition of states $|J,M\rangle$ in terms of states $|m_1 , m_2 \rangle$, for $j_1 = j_2 = 3/2$. Specifically, $|J,M\rangle = \sum_{m_1} S^J_{m_1, M - m_1} |m_1 , M - m_1 \rangle$. We use the Condon--Shortley phase convention.}
    \label{tab:my_label}
\end{table}

We now list the resulting spin--orbital wave functions. The singlet ground state is
\begin{align}
\big|  \mathbb{S} \big>  & =  \Big( \big|\Psi^S_{1H, 2H}\big> + \big|\Psi^S_{1L, 2L}\big> \Big) \otimes  \big|0,0\big>   \nonumber \\
& \quad   +   \Big( \big|\Psi^S_{1H, 2H}\big>  - \big|\Psi^S_{1L, 2L}\big>    \Big) \otimes  \big| 2,0\big>      
\nonumber \\
& \quad +  \frac{1}{\sqrt{2}} \,   \Big(  \big|\Psi^S_{1L, 2H}\big> +\big|\Psi^S_{1H, 2L}\big>  \Big) \! \otimes \! \Big(  \big|2,2\big>   + \big|2,-2\big>    \Big)  \nonumber \\
 & \quad    +  \frac{1}{\sqrt{2}} \,  \Big(   \big|\Psi^A_{1L, 2H}\big>  -  \big| \Psi^A_{1H, 2L}\big> \Big) \! \otimes \! \Big(    \big|3,2\big>   - \big|3,-2\big>     \Big)  \,,  
 \label{Singlet Bz}
\end{align}
and the triplet states are
\begin{align}
\big|  T_{z, +} \big>    & =        
 \sqrt{2} \,   \big|\Psi^A_{1H, 2H}\big>    \otimes   \big|3,3\big>  \nonumber \\
 & \quad +  \sqrt{2}  \,    \big|\Psi^A_{1L, 2L}\big>       \otimes        \left(  \sqrt{ \frac{2}{5} } \,   \big|1,-1\big> +  \sqrt{ \frac{3}{5} } \,   \big|3,-1\big>    \right)    
\nonumber \\ 
& \quad +    \Big( \big|\Psi^A_{1L, 2H}\big> + \big|\Psi^A_{1H, 2L} \big>    \Big)  \otimes          \left( \sqrt{\frac{3}{5}} \, \big|1,1\big> \right. \nonumber \\
& \quad \left. - \, \sqrt{\frac{2}{5}}  \, \big|3,1\big> \! \right)                
 \! +    \Big(   \big|\Psi^S_{1L, 2H} \big> -  \big| \Psi^S_{1H, 2L}   \big> \Big)  \! \otimes  \! \big|2,1\big>        \,,     
\label{Triplet + Bz}
\end{align}
\begin{align}
\big|  T_{z, 0} \big>    & =           \frac{1}{\sqrt{ 2}} \, \Big(  \big| \Psi^A_{1L, 2H}\big> \!+ \! \big| \Psi^A_{1H, 2L} \big>    \Big)      \otimes   \Big( \big| 3,2 \big> \! + \! \big|3,-2 \big> \Big)              
\nonumber \\ 
& \quad +    \frac{1}{\sqrt{ 2} }  \Big(  \big| \Psi^S_{1L, 2H} \big> \!- \! \big|  \Psi^S_{1H, 2L} \big>    \Big) \! \otimes         \! \Big(  \big| 2,2 \big> \! - \!\big| 2,-2 \big>   \Big)                  
\nonumber \\ 
& \quad +   \sqrt{\frac{1}{5}} \Big[    \Big(     3 \big| \Psi^A_{1H, 2H} \big>      -  \big| \Psi^A_{1L, 2L} \big> \Big) \otimes \big| 1,0 \big>    \nonumber \\
& \quad +   \Big(       \big| \Psi^A_{1H, 2H} \big>     +  3    \big| \Psi^A_{1L, 2L} \big>    \Big) \otimes \big| 3,0 \big> \Big]   \,,
\label{Triplet 0 Bz}
\end{align}
and
\begin{align}
\big|  T_{z, -} \big>    & =  \sqrt{2} \, \big| \Psi^A_{1H, 2H}  \big> \otimes  \big| 3,-3  \big>         \nonumber \\
& \quad     - \sqrt{2} \,     \big| \Psi^A_{1L, 2L}  \big> \otimes   \left( \sqrt{ \frac{2}{5} }   \, \big| 1,1  \big> + \sqrt{ \frac{3}{5} }   \, \big| 3,1  \big> \right) 
\nonumber \\ 
& \quad -   \Big(  \big| \Psi^A_{1L, 2H}  \big> + \big|\Psi^A_{1H, 2L}  \big>      \Big)    \otimes      \left( \sqrt{\frac{3}{5}} \, \big| 1,-1  \big>  \right. \nonumber \\
& \quad  \left. - \, \sqrt{\frac{2}{5}} \, \big| 3,-1  \big>   \right)  \nonumber \\
& \quad     +     \Big(   \big| \Psi^S_{1L, 2H}  \big>  - \big| \Psi^S_{1H, 2L}  \big>     \Big)   \otimes      \big|2,-1  \big>           \,.
\label{Triplet - Bz}
\end{align}

The good agreement between this analytical theory and the numerical results is discussed in Appendix \ref{App: numerical 2 holes vs Hubbard}.

One of the interesting features of the two--hole states is that they are not eigenstates of the total angular momentum operators $\hat{\boldsymbol{J}}^2$ and $\hat{J}_z$. Therefore, as seen from the expressions \eqref{Singlet Bz}, \eqref{Triplet + Bz}, \eqref{Triplet 0 Bz} and \eqref{Triplet - Bz}, they are linear combinations of two--hole spinors $\big| J, M \big>$, with various values of $J$ and $M$. Moreover, the spinors contributing to each state do not all have the same symmetry under exchange of the two particles, and consequently the same holds true for the orbitals. Therefore, two--hole eigenstates cannot be factorized as products of orbital and spin parts. This mixing is a peculiar feature of the four--band system, which is strikingly different from more common one--band systems. However, we have found numerically that the weights of the antisymmetric orbitals in the singlet and of the symmetric orbitals in the triplet are small (details are given in Section \ref{subsec: twoholenum} and Appendix \ref{App: numerical 2 holes vs Hubbard}). The smallness of these terms, in the systems that we have considered, can be interpreted as a consequence of the small degree of entanglement between orbital and spin degrees of freedom, which we have already demonstrated via the calculation of the linear entropies associated with the spin reduced density matrix. We discuss this connection in Section \ref{Sec: spin rep}. 

We also notice that, within the present model, non--zero light--hole components are necessary for the even--odd $J$ mixing: if we were to set $\left| \psi_{sL} \right> \rightarrow 0$, the two--hole states would reduce to
\begin{align}
& \big|  \mathbb{S} \big>   \rightarrow  \sqrt{2} \, \big|\Psi^S_{1H, 2H}\big> \otimes \frac{1}{ \sqrt{2} }  \Big( \big|0,0\big>   +  \big| 2,0\big> \Big)     
   \,,   \nonumber \\ 
& \big|  T_{z, +} \big>    \rightarrow       
 \sqrt{2} \,   \big| \Psi^A_{1H, 2H}\big>    \otimes  \big|3,3\big> \,,  \nonumber \\    
& \big|  T_{z, 0} \big>     \rightarrow      \sqrt{2} \,   \big|    \Psi^A_{1H, 2H}\big>   \otimes \sqrt{\frac{1}{10}}  \Big(     3      \big|1,0\big>      +  \big|3,0\big>      \Big)   \,, \nonumber \\
& \big|  T_{z, -} \big>     \rightarrow  \sqrt{2} \, \big|\Psi^A_{1H, 2H}\big> \otimes  \big| 3,-3\big>        \,. 
\label{single-band limit}
\end{align}  
These are in agreement with the single--band limits discussed in Section \ref{subsec: twoholenum}.

\section{Pseudospin representation of the one-- and two--hole states}
\label{Sec: spin rep}

Since the lowest single--hole eigenstates are characterized by a limited amount of mixing between orbital and spin components (see Sec. \ref{Sec: Two}), they can be approximately described within an effective spin picture, and represented as combinations of spinors $\big| J, M \big>$ only.

\subsection{Single--hole states}

In particular, single--hole states $|\psi_\alpha\rangle$, with $\alpha \in \lbrace 1,2,3,4 \rbrace$, are here characterized by a dominant heavy--hole component $m=\pm 3/2$ and a significant contribution from the light--hole $m=\mp 1/2$ (see Table \ref{tab:single_hole} in Subsec.~\ref{subsec:shs}). Besides, both single-- and two--hole lowest--energy states display a limited amount of entanglement between spin and orbital degrees of freedom. As a reference for the actual reduced spin states, we consider hereafter the limiting case where such entanglement is exactly zero, due to a perfect overlap between the orbital states corresponding to different bands:
\begin{align}
 \big|\psi_{s, L}\big> \approx - r   \big|\psi_{s, H}\big> \,, \,\,\,
  \big|\psi_{s, H}\big> \approx     \big|\psi^*_{s, H}\big> \,, \,\,\,
  \big|\psi_{s, L}\big> \approx     \big|\psi^*_{s, L}\big> \,,
\label{freezing orbital}
\end{align}
with $s \in \lbrace 1, 2 \rbrace$. One has $r^2 = p_{i,\mp 1/2}/p_{i,\pm 3/2}$; the particular case of $r=0$ reproduces the single--band limit [see Eqs.~\eqref{single-band limit}]. The approximation \eqref{freezing orbital}, for arbitrary $r$, amounts to freezing the orbital degrees of freedom, since it implies that there is only one independent orbital function for each dot; single--hole states are then factorized as in Eq.~\eqref{factorized state}.

This allows us to introduce a pseudospin--$1/2$ representation for the single--hole states localized in each dot, and express them exclusively in terms of the eigenstates $|m\rangle$ of $j_z$:  
\begin{align}
& \big| \uparrow \big>_s \equiv  \frac{1}{\sqrt{1+r^2}} \Big( \big| \tfrac{3}{2} \big>_s - r \big| \! - \tfrac{1}{2} \big>_s \Big) \,, \nonumber\\
& \big| \downarrow \big>_s \equiv \frac{1}{\sqrt{1+r^2}} \Big( \big| \! -\tfrac{3}{2} \big>_s - r \big| \tfrac{1}{2} \big>_s \Big) \,.
\label{effective spin states single dot}
\end{align}
Each spinor for site $s$ incorporates the heavy--hole orbital wave function for dot $s$, namely, $\psi_{s, H}(\boldsymbol{r})$.

\subsection{Two--hole states}

The four lowest two--holes states in the pseudospin--$1/2$ representation are obtained from Eqs.~\eqref{Singlet Bz}, \eqref{Triplet + Bz}, \eqref{Triplet 0 Bz} and \eqref{Triplet - Bz} by applying the approximations \eqref{freezing orbital} to the single--hole orbitals that constitute the two--hole orbitals. Equivalently, they can be obtained by writing the singlet and triplet states in terms of two pseudospin--$1/2$ single--hole states as follows,
\begin{align} 
 & \big| S^{\rm ps} \big>   =   \frac{1}{\sqrt{2}} \Big( \big| \downarrow \big>_1 \big|\uparrow \big>_2 - \big| \uparrow \big>_1  \big| \downarrow \big>_2  \Big)   \,, \nonumber \\
 &    \big| T^{\rm ps}_+ \big>   =   \big| \uparrow \big>_1 \big| \uparrow \big>_2   \,, \nonumber \\  
&    \big| T^{\rm ps}_0 \big>   =  \frac{1}{\sqrt{2}} \Big( \big| \downarrow \big>_1 \big|\uparrow \big>_2 + \big| \uparrow \big>_1  \big| \downarrow \big>_2 \Big)   \,, \nonumber \\
&    \big| T^{\rm ps}_- \big>   =    \big| \downarrow \big>_1 \big| \downarrow \big>_2   \,,
\end{align} 
and replacing states $\big| \uparrow  \big>_s$ and $\big| \downarrow  \big>_s$ with the expressions given in Eqs.~\eqref{effective spin states single dot}. One obtains the following expansions in the $\big| J, M \big>$ basis:
\begin{align}
  \big| S^{\rm ps} \big>  & =    \frac{1}{ 1 + r^2 } \Bigg[  \frac{1}{\sqrt{2}} \Big(1+r^2 \Big)  \big| 0,0 \big>          
\nonumber \\
& \quad  +  \frac{1}{\sqrt{2}} \Big(1-r^2 \Big) \big| 2,0 \big>  -      r \Big(  \big| 2,2 \big>   + \big| 2,-2 \big>    \Big) \Bigg]  \,,  
 \label{Singlet Bz - spin model approximation}
\end{align} 
\begin{align}
 \big|  T^{\rm ps}_+ \big>    & =
    \frac{1}{   1 + r^2  }     \Bigg[     \big| 3,3 \big>    +              r^2        \left(  \sqrt{ \frac{2}{5} } \,   \big| 1,-1 \big> +  \sqrt{ \frac{3}{5} } \,   \big| 3,-1 \big>    \right)    
\nonumber \\ 
& \quad - \sqrt{2} r           \left( \sqrt{\frac{3}{5}} \, \big|1,1\big>   -   \sqrt{\frac{2}{5}} \, \big|3,1\big> \right)  \Bigg]              
          \,,     
\label{Triplet + Bz - spin model approximation}
\end{align}
\begin{align}
  \big|  T_0^{\rm ps} \big>    & = \frac{1}{   1 + r^2  }   \Bigg[ -   r  \, \Big( \big| 3,2 \big> + \big| 3,-2 \big> \Big)              
\nonumber \\ 
& \quad +      \sqrt{\frac{1}{10}}   \Big(     3     -  r^2     \Big) \, \big| 1,0 \big>     +   \sqrt{\frac{1}{10}}   \Big(   1    +  3    r^2    \Big) \, \big| 3,0 \big>      \Bigg]   \,,
\label{Triplet 0 Bz - spin model approximation}
\end{align}
and
\begin{align}
  \big|  T_-^{\rm ps} \big>    & = \frac{1}{   1 + r^2  }  \Bigg[   \big| 3,-3 \big>             -   r^2 \,          \left( \sqrt{ \frac{2}{5} }   \, \big| 1,1 \big>   + \sqrt{ \frac{3}{5} }   \, \big| 3,1 \big> \right) 
\nonumber \\ 
& \quad + \sqrt{2} r \left( \sqrt{\frac{3}{5}} \, \big| 1,-1 \big>    -   \sqrt{\frac{2}{5}} \, \big| 3,-1 \big>   \right)   \Bigg] \,.
\label{Triplet - Bz - spin model approximation}
\end{align}
It is intended that the two--hole spinors $\big| J, M \big>$ incorporate the information about two--hole orbitals; namely, spinors with even $J$ (antisymmetric) incorporate $\Psi^{S}_{1H, 2H}(\boldsymbol{r}_1, \boldsymbol{r}_2)$, while spinors with odd $J$ (symmetric) incorporate $\Psi^{A}_{1H, 2H}(\boldsymbol{r}_1, \boldsymbol{r}_2)$.

By comparing the two--hole spinors entering Eqs.~\eqref{Singlet Bz - spin model approximation}, \eqref{Triplet + Bz - spin model approximation}, \eqref{Triplet 0 Bz - spin model approximation} and \eqref{Triplet - Bz - spin model approximation} with those entering the Hubbard states, we notice that the difference of antisymmetric orbitals which multiplies the $J=3$ spinors in the Hubbard singlet [Eq.~\eqref{Singlet Bz}] cancels out as a consequence of approximation \eqref{freezing orbital}, just as the differences between symmetric orbitals multiplying the $J=2$ spinors in the triplet states [Eqs.~\eqref{Triplet + Bz}, \eqref{Triplet 0 Bz} and \eqref{Triplet - Bz}]. The singlet now contains only spinors with even $J$ (antisymmetric), while the triplet contains only spinors with odd $J$ (symmetric); accordingly, all components of the singlet have the symmetric orbital $\Psi^{S}_{1H, 2H}(\boldsymbol{r}_1, \boldsymbol{r}_2) $, while all components of each triplet state have the antisymmetric orbital $\Psi^{A}_{1H, 2H}(\boldsymbol{r}_1, \boldsymbol{r}_2) $.

Therefore, within the Hubbard model the source of even/odd $J$ mixing is the difference in the spatial dependence (non--parallelism) of the heavy--hole and light--hole orbitals, as we have just seen that, when this difference is removed, the mixing disappears. However, the two--hole states in the pseudospin--$1/2$ formalism are still not eigenstates of $\hat{\boldsymbol{J}}^2$ and $\hat{J}_z$, an effect of spin--orbit coupling. At the single--hole level, this can be traced to the fact that it is not possible to choose a basis of single--hole states which are eigenstates of $\hat{J}_z$ [see Eqs.~\eqref{effective spin states single dot}].

\section{Conclusions}
\label{Sec: Conclusions}

In conclusion, we have applied different theoretical approaches to investigate the properties of single-- and two--hole states in prototypical coupled Si and Ge quantum dots. These states are comprehensively modeled within a 6--band $\boldsymbol{k} \cdot \boldsymbol{p}$ and Configuration--Interaction approach, from which we extract the band mixing, the weight of the relevant orbitals, and the reduced spin states corresponding to the lowest two--hole eigenstates. In particular, we propose the use of entanglement measures (such as the linear entropy) to achieve a deeper characterization of the band mixing, beyond what is allowed by the study of band--occupation probabilities. The numerical results are used to define the range of validity of effective representations, provided by a reduced spin model and by a generalized Hubbard model, which in turn allows for an analytic derivation of the spin states. The lowest two--hole eigenstates display both analogies and differences with respect to the singlet and triplet states obtained in two--electron (one--band) systems. In particular, the singlet ground state and the triplet excited states are predominantly antisymmetric and symmetric with respect to spin exchange, respectively, as in the electron case. However, unlike in that case, they also display a strong $J$--mixing, even in the absence of band mixing. The presence of light--hole components additionally results in $M$--mixing within spin subspaces having the same permutation symmetry, as well as in small contributions from spin subspaces with opposite symmetry with respect to the dominant one.

\acknowledgments
The authors acknowledge financial support from the European Commission through the project IQubits (Call: H2020--FETOPEN--2018--2019--2020--01, Project ID: 829005).

\appendix
\section{Comparison between the Hubbard model and the numerical results}
\label{App: numerical 2 holes vs Hubbard}

In order to compare the predictions of the Hubbard model with our numerical results, we consider the weights of the $(J, M)$ components of the four lowest two--hole states, i.e., $p_{k, J, M}$, with $k \in \lbrace 1,2,3,4 \rbrace$ corresponding to states $\lbrace \mathbb{S}, T_{z,+}, T_{z,0}, T_{z,-} \rbrace$. From the expressions of two--hole states, Eqs.~\eqref{Singlet Bz}, \eqref{Triplet + Bz}, \eqref{Triplet 0 Bz}, and \eqref{Triplet - Bz}, we see that the Hubbard model predicts precise relationships between the weights of certain $(J,M)$ components in different states, or within the same state. These relationships, which hold independently of the specific form of the single--hole wave functions, are: 
\begin{align}
& p_{1,2,2} = p_{1,2,-2} \,, \quad p_{1,3,2} = p_{1,3,-2} \,, \nonumber \\
& p_{2,3,3} = p_{4,3,-3} \,, \quad p_{2,1,-1} = p_{4,1,1} \,, \quad p_{2,3,-1} = p_{4,3,1} \,, \nonumber \\
& p_{2,1,1} = p_{4,1,-1} \,, \quad p_{2,3,1} = p_{4,3,-1} \,, \quad p_{2,2,1} = p_{4,2,-1} \,;
\label{set rel 1}
\end{align}
\begin{align}
& p_{3,1,0} + p_{3,3,0} - p_{2,3,3} - \frac{5}{2} p_{2,1,-1} \nonumber \\
& = p_{3,1,0} + p_{3,3,0} - p_{4,3,-3} - \frac{5}{2} p_{4,1,1} = 0 \,; 
\label{set rel 2}
\end{align}
\begin{align}
& \frac{p_{2,1,1}}{p_{2,3,1}} = \frac{p_{2,3,-1}}{p_{2,1,-1}} = \frac{p_{4,1,-1}}{p_{4,3,-1}} = \frac{p_{4,3,1}}{p_{4,1,1}} = \frac{3}{2} \,, \nonumber \\
& \frac{p_{2,3,1}}{p_{3,3,2}}   = \frac{p_{4,3,-1}}{p_{3,3,2}}   = \frac{4}{5} \,, \nonumber \\
& \frac{p_{2,2,1}}{p_{3,2,2}}   = \frac{p_{4,2,-1}}{p_{3,2,2}}   = 2  \,.
\label{set rel 3}
\end{align} 
Moreover, the model predicts certain quantities $p_{k, J, M}$ to be zero, since each of the states $\lbrace \mathbb{S}, T_{z,+}, T_{z,0}, T_{z,-} \rbrace$ only involves 6 spinors $\big| J,M \big>$ out of 16.

To estimate the accuracy of the model, we compare its predictions with the two--hole calculations performed for silicon, in the exemplary cases of $a \in \lbrace 5, 10, 14 \rbrace$ nm. The results of the comparison apply similarly to all other calculations that we have performed, for both silicon and germanium. 

We find that all identities \eqref{set rel 1} and linear combinations \eqref{set rel 2} are satisfied within an error $\lesssim 10^{-4}$, which is comparable to the estimated accuracy of the numerical calculations. The ratios given by Eqs.~\eqref{set rel 3} are shown in Table \ref{tab:Hubbard comparison}: the second column provides the model prediction, while columns 3--5 provide the numerical results for the three cases that we are considering. Some of the weights ($p_{2,3,-1}$, $p_{2,1,-1}$, $p_{4,3,1}$ and $p_{4,1,1}$) involved in Eqs.~\eqref{set rel 3} are found to be $\approx 10^{-4}$, which is below the estimated accuracy of the numerical calculations; therefore, computing their ratios is not significant, and we omit those from our analysis. This smallness can be interpreted within the Hubbard model by assuming that, in the systems considered in our calculations, the orbital $\big| \Psi^A_{1L, 2L} \big> $ has a small amplitude [see Eqs.~\eqref{Triplet + Bz} and \eqref{Triplet - Bz}]. In all significant cases, the model accurately accounts for the numerical results.

\begin{table} 
\begin{tabular}{ |c||c|c|c|c| } 
 \hline
 quantity & model & $a = 5$ nm & $a = 10$ nm & $a = 14$ nm \\ 
 \hline
 $p_{2,1,1} / p_{2,3,1}$ & $1.5$ & $1.515$ & $1.500$ & $1.500$ \\ 
 $p_{4,1,-1} / p_{4,3,-1}$ & $1.5$ & $1.488$ & $1.500$ & $1.500$ \\ 
 \hline   
 $p_{2,3,1} / p_{3,3,2}$ & $0.8$ & $0.795$ & $0.800$ & $0.809$ \\
 $p_{4,3,-1} / p_{3,3,2}$ & $0.8$ & $0.803$ & $0.800$ & $0.809$ \\
 \hline
 $p_{2,2,1} / p_{3,2,2}$ & $2$ & $1.987$ & $2.000$ & $2.000$ \\
 $p_{4,2,-1} / p_{3,2,2}$ & $2$ & $1.987$ & $2.000$ & $2.000$ \\
 \hline
\end{tabular} 
 \caption{Comparison between the predictions of the Hubbard model and the numerical results.}
    \label{tab:Hubbard comparison}
\end{table}

We find that the states obtained from the numerical calculations include very small, but non--zero, contributions from some $(J,M)$ components for which $p_{k,J,M} = 0$ according to the model. We call $h_k$ the {\it sum} of the weights of these $(J,M)$ components beyond the Hubbard model for state $k$. For the singlet, we obtain $h_1 \approx \lbrace 6, 5, 5\rbrace \times 10^{-3}$ for the three considered values of $a$, respectively. The weights of the individual terms beyond Hubbard are all $\lesssim 10^{-4}$, except for $p_{1,2,1} = p_{1,2,-1} = 1.3 \times 10^{-3}$ in all the three cases. For the triplet, we obtain $h_2 \approx h_3 \approx h_4 \approx \lbrace 8, 6, 5 \rbrace \times 10^{-3}$ for the three values of $a$, respectively. The weights of the individual terms beyond Hubbard are all $\lesssim 10^{-4}$, except for $p_{2,3,2} = p_{2,2,2} = p_{4,3,-2} = p_{4,2,-2}$, which is equal to $\approx \lbrace 3.5, 2.7, 2.5 \rbrace \times 10^{-3}$ for the three values of $a$, respectively.

One of the predictions of the Hubbard model in this four--band scenario is that both symmetric and antisymmetric two--hole orbitals contribute to the singlet and triplet states. The weight of the antisymmetric orbitals in the singlet is given by the quantities $p_{1,3,2} = p_{1,3,-2}$, while the weight of the symmetric orbitals in the triplet is given by the quantities $p_{2,2,1} = p_{4,2,-1}$ and $p_{3,2,2} = p_{3,2,-2}$ [compare with Eqs.~\eqref{Singlet Bz}, \eqref{Triplet + Bz}, \eqref{Triplet 0 Bz}, \eqref{Triplet - Bz}]. We report them in Table \ref{tab:odd ones} for the same three cases considered above. In all cases, it is seen that the orbitals with minority symmetry have a total weight of the order of $\approx 10^{-2}$. The connection between this smallness and the small degree of spin--orbital entanglement is discussed in Section \ref{Sec: spin rep}.

\begin{table} 
\begin{tabular}{ |c||c|c|c| } 
 \hline
 quantity & $a = 5$ nm & $a = 10$ nm & $a = 14$ nm \\ 
 \hline
 $p_{1,3,2} = p_{1,3,-2}$ & $4 \times 10^{-3}$ & $6 \times 10^{-3}$ & $5 \times 10^{-3}$ \\
 $p_{2,2,1} = p_{4,2,-1}$ & $16 \times 10^{-3}$ & $12 \times 10^{-3}$ & $10 \times 10^{-3}$ \\ 
 $p_{3,2,2} = p_{3,2,-2}$ & $8 \times 10^{-3}$ & $6 \times 10^{-3}$ & $5 \times 10^{-3}$ \\ 
 \hline
\end{tabular} 
 \caption{Weights of the antisymmetric orbitals in the singlet and of the symmetric orbitals in the triplet.}
    \label{tab:odd ones}
\end{table}

\end{document}